\documentstyle[12pt]{article}                     
\documentstyle{fleqn}                             

\begin{document}                                  
\begin{flushright}                                
OU-HET 242                                        
\end{flushright}                                  
\vskip1.5cm                                       
\begin{center}                                    
{ \large                                          
{\bf                                              
Generalized screening theorem 
for  Higgs  decay processes 
\\                                                
 in the two-doublet model                         
}}                                                
\end{center}                                      
\vskip1.5cm                                       
\begin{center}                                    
{\large Shinya Kanemura\footnote                  
{Address after April 1st, 1996: Theory Group, KEK,
Tsukuba, Ibaraki, 305, Japan},                    
 Takahiro Kubota                                  
and Hide-Aki Tohyama}                             
\end{center}                                      
\vskip1.5cm                                       
\begin{center}                                    
{\it Department of Physics,   Osaka University,   
\linebreak                                        
Toyonaka, Osaka 560, Japan}                       
\end{center}                                      
\vskip0.5cm                                       
\vfill\eject                                      
\begin{center}                                    
{\bf Abstract}                                    
\end{center}                                      
The radiative corrections to the decay processes  
of the neutral ($CP$-even) Higgs boson ($H$) into 
a longitudinal gauge boson pair, {\it i.e.},      
$H \rightarrow Z_{L}Z_{L}$ and $H \rightarrow     
W_{L}^{+}W_{L}^{-}$ are analyzed in the two-Higgs 
doublet model by assuming that all of the 
Higgs boson masses 
are much greater than the $W$ and $Z$ bosons'.  
These calculations are motivated to see if one could 
see potentially large virtual 
effects to these decay rates due to the charged 
and $CP$-odd neutral Higgs boson masses ($m_{G}$ and 
$m_{A}$, respectively) which are supposed to be 
larger than $m_{H}$.
It is pointed out that, although  the 
radiative corrections to the decay width 
$\Gamma (H\rightarrow W_{L}^{+}W_{L}^{-})$
depend  sensitively in general on 
$m_{G}$ and $m_{A}$,  there occurs a screening 
effect,  
{\it i.e.,} cancellation in  leading terms once 
we set 
$m_{G}=m_{A}$,  so that the radiative corrections 
 tend to be minimized. It is also pointed out 
that the decay rate 
$\Gamma (H\rightarrow Z_{L}Z_{L})$ is fairly
insensitive to the other heavier Higgs masses and
 is possibly a good measuring tool 
 of the Higgs mixing angle.
The mechanism of these  screening phenomena 
in the Higgs decays is  
explained on the basis of a new screening theorem, 
which we postulate with reference to  the custodial 
symmetry in  the Higgs potential.

\pagebreak
\noindent
{\large
{\bf 1. Introduction}}
\vskip0.5cm

It is becoming more and more important to investigate 
non-decoupling effects of heavy particles in the low 
energy observables, since  
the available energy of future  accelerators will 
increase only little by little. 
One of the examples of low energy manifestations of 
heavy particles  has been  given by the top quark, 
whose mass was surmised before its discovery [1,2] by  
detailed analyses  of the electroweak data.

Now that the existence of the top quark has been 
confirmed, the next pressing experimental task  is the 
discovery of the Higgs boson. (For a review, 
see Ref. [3].)  It has been known for 
long time that the Higgs boson in the standard model is 
very elusive: indirect signatures of the Higgs boson 
appear in the low energy data on the oblique-type 
radiative corrections at most in  logarithmic 
terms at the one loop level. This fact is often referred 
to as Veltman's screening theorem [4]. 
Perhaps more precise 
electroweak  data would change the situation and the 
signatures of the standard model Higgs boson might be 
just around the corner.
In any case, once the existence of the standard model 
Higgs boson would be established, we should be   still
looking after new unknown heavy particles even further, 
via radiative corrections in order  to probe 
what lies beyond the standard model. 

In our previous publication [5], we have investigated 
the possibility that one could get hold of 
signatures of unknown 
heavy particles through radiative corrections, 
supposing that the standard model 
Higgs boson $H$ has been discovered.   
As an example we adopted a two-Higgs doublet model, 
the most conservative extension 
 of the standard model. Here   
the Higgs sectors consist of, besides the neutral 
$H$ boson,  another ($CP$-even) neutral Higgs boson $h$, 
($CP$-odd) neutral Higgs boson 
$A$, and the charged Higgs boson $G^{\pm}$.
Assuming that their masses $m_{H}$, $m_{h}$, $m_{A}$ 
and $m_{G}$ are all much greater than $M_{W}$, 
we computed the radiative corrections to the decay width
$\Gamma (H\rightarrow W_{L}^{+}W_{L}^{-})$ 
as a function of $m_{h}$, $m_{A}$, and $m_{G}$. Here  
the subscript $L$ denotes the longitudinal polarization .
The preliminary numerical calculations in Ref. [5] show 
that the magnitude of radiative corrections depends  
rather sensitively on the choice of $m_{G}$ and $m_{A}$.
It is also suggested implicitly  that, if $m_{G}=m_{A}$, 
then  the radiative corrections are minimized. Note 
in this connection that the Higgs potential in the 
two-doublet model respects an 
$SU(2)_{L}\times SU(2)_{R}$
 symmetry [6], if we put  $m_{G}=m_{A}$.
The existence of the additional global 
$SU(2)_{R}$ symmetry would lead to isospin symmetry 
$SU(2)_{V}={\rm diag}[SU(2)_{L}\times SU(2)_{R}]$ 
after spontaneous symmetry breaking. This isospin 
symmetry is often called custodial symmetry in
 literatures.

The purpose of the present paper is 
two-fold. The first one is to extend our 
previous analyses to another decay mode
$H\rightarrow Z_{L}Z_{L}$ 
and to compare its dependences on $m_{G}$, 
$m_{A}$,  $m_{h}$ and Higgs mixing angles ($\alpha $ 
and $\beta $) with  those in 
$H\rightarrow W_{L}^{+}W_{L}^{-}$.
 We would like to establish 
 the following two peculiar facts as to the radiative
 corrections by numerical calculations on computer.
\begin{description}
\item{(1)}
Although the radiative corrections to the decay 
$H\rightarrow W_{L}^{+}W_{L}^{-}$ is in general 
sensitive to the choice of $m_{G}$ and $m_{A}$, 
they  tend to be minimized if the mass parameters 
are such  that the custodial symmetry is respected, 
{\it i.e.},  $m_{G}=m_{A}$.
\item{(2)}
The radiative corrections to the decay 
$H\rightarrow Z_{L}Z_{L}$ 
are  insensitive to the choice of the mass 
parameters. It is only sensitive to the 
Higgs mixing angles.
\end{description}
\noindent
The second purpose of the present work 
is to give theoretical explanations 
to (1) and (2) without recourse to numerical methods. 
We will postulate  a new screening ``theorem" for the
 Higgs decay vertices, by  which we are able to 
acquire a clear grasp of 
$m_{G}$-, $m_{A}$- and $m_{h}$-dependences of 
the decay widths 
$\Gamma (H\rightarrow W_{L}^{+}W_{L}^{-})$ and 
$\Gamma (H\rightarrow Z_{L}Z_{L})$.

The screening theorem for the Higgs vertices 
that we just mentioned 
is in close analogy with the celebrated 
Veltman's theorem [4], 
which may be applied, in contrast,  to
the oblique-type radiative corrections.  
The statement in  (1) referring to the custodial 
 symmetry reminds 
us of the detailed study of the 
Veltman's theorem in the standard model 
by Einhorn and Wudka [7].   They argued that the 
radiative corrections to the gauge boson propagators 
may be  classified into several types. On the basis of 
the custodial symmetry in the weak  
 $U(1)_{Y}$-coupling limit, 
they claimed that the Higgs mass 
dependence at the $L$-th 
loop is at most $(m_{H}^{2})^{L-1}$ rather than 
$(m_{H}^{2})^{L}$. 
The statement (1) indicates that there is a similar  
situation in the two-Higgs doublet model as well, 
in  radiative corrections to the (non-oblique type) 
decay process if the custodial symmetry is respected.  
We will make this fact crystalline 
 in the form of a  new screening  theorem.
Although we will not go  so far as to 
give as general an argument as Einhorn and Wudka's 
including  all order corrections, 
the similarity lying between Veltman's theorem and our 
counterpart provides a strong evidence in favor 
of the validity of our version beyond the one-loop level.

Throughout the present paper we will often refer to 
 and make use of some of the formulae given in Ref. [5] 
with respect to the decay 
$H\rightarrow W_{L}^{+}W_{L}^{-}$.
In Ref. [5], in passing,  the effect of the top 
quark mass was not taken into consideration. 
Since the top quark mass 
might  produce non-negligible effects  to 
the decay [8], we will improve our previous analysis 
by including quark loops as well. 
Partial lists of  earlier works on  these decay processes 
in the standard model and  supersymmetric models
are given in Ref. [9-12]. A brief summary of the 
present work is found in Ref. [13].

The present paper is organized  as follows. Our Higgs 
potential is specified in Sec. 2, thereby explaining the 
connection between the custodial symmetry and the mass 
degeneracy $m_{G}=m_{A}$. We will make an extensive  use 
of the equivalence theorem [14-17] at the loop level. 
Sec. 3 is devoted to clarification of an issue that arises 
if one uses  the equivalence theorem at higher 
orders [17]. 
One loop calculations of the decay processes, which will 
be presented in Sec. 5,  are rendered  complicated, 
 because of the mixing 
between those of  same quantum 
numbers. The method of renormalization of the 
Higgs mixing angles and the
wave function renormalization constant matrices are 
explained at length in Sec. 4.
Our numerical calculation is presented in Sec. 6.  
Our new screening theorem is given in Sec. 7 which, 
we believe, will lay cornerstones for clear understanding 
of  the  qualitative features of the numerical 
calculations in Sec. 6. Sec. 8 is devoted to summary and 
discussions.  Some details of our calculations are  
 relegated to several Appendices.

\vskip2cm
\noindent 
{\large {\bf 2. The Higgs potential and the custodial 
symmetry }}
\vskip0.5cm 

Before launching into the details of our calculations, 
we have to specify our Higgs potential of  the 
two Higgs doublets,   $\Phi _{1}$ and 
$\Phi _{2}$ with $Y=1$. The criterion of determining  
 the Higgs potential is the natural suppression of the 
flavor-changing neutral current [18] 
and it is often assumed 
that the discrete  symmetry 
$\Phi _{2} \rightarrow -\Phi _{2}$ is respected except 
for soft terms, namely, 
\begin{eqnarray}
V(\Phi _{1}, \Phi _{2})&=& 
-\mu _{1}^{2}\mid \Phi _{1}\mid  ^{2}
-\mu _{2}^{2}\mid \Phi _{2}\mid ^{2}
-(\mu _{12}^{2}\Phi _{1}^{\dag}\Phi _{2}+\mu _{12}^{2*}
\Phi _{2}^{\dag}\Phi _{1})       \nonumber \\
& &
+\lambda _{1}\mid \Phi _{1}\mid ^{4}
+\lambda _{2}\mid \Phi _{2}
\mid ^{4}
+ \lambda _{3}\mid \Phi _{1}\mid  ^{2}\mid  \Phi _{2}\mid 
^{2}  \nonumber \\
& &+ \lambda _{4}({\rm Re}\Phi _{1}^{\dag}\Phi _{2})^{2}
+\lambda _{5}({\rm Im }\Phi _{1}^{\dag}\Phi _{2})^{2}.
\end{eqnarray}
This potential is general enough to encompass 
supersymmetric models and 
the soft  term 
$(\mu _{12}^{2}\Phi _{1}^{\dag}\Phi _{2}+\mu _{12}^{2*}
\Phi _{2}^{\dag}\Phi _{1})$
 plays an important role there. 
The complex phase of $\mu _{12}$   may also be  
instrumental  to the idea of baryogenesis 
at the electro-weak scale [19]. The existence of this term, 
however, makes our analysis too  complicated 
to get an insight into radiative corrections. 
We will therefore set $\mu _{12}=0$ throughout our 
calculations. Perhaps it is worth mentioning 
as another  excuse 
for setting $\mu _{12}=0$ that, since we are interested 
in the non-decoupling effects caused by strong quartic 
couplings (times $v^{2}\approx (246 {\rm GeV})^{2}$), 
the mass scale $\mu _{12}$ as opposed to 
$v$ does not have direct relevance  to 
what we are concerned with and 
may be neglected at the first step.

From the viewpoint of the custodial symmetry, it is 
convenient to introduce a 
$2\times 2$ matrix notation of the Higgs 
fields, {\it i.e.}, 
${\bf \Phi }_{i}=(i\tau _{2}\Phi _{i}^{*},  \Phi _{i}).$
In terms of this notation, our Higgs potential is 
expressed as  
\begin{eqnarray}
V(\Phi _{1}, \Phi _{2})&=&
-\frac{1}{2}\mu _{1}^{2}{\rm tr}({\bf 
\Phi }_{1}^{\dag}{\bf \Phi }_{1})
-\frac{1}{2}\mu _{2}^{2}{\rm tr}({\bf 
\Phi }_{2}^{\dag}{\bf \Phi }_{2})    \nonumber \\
& &+\frac{1}{4}\lambda _{1}\left \{{\rm tr}({\bf 
\Phi }_{1}^{\dag}{\bf \Phi }_{1})\right \}^{2}
+\frac{1}{4}\lambda _{2}\left \{{\rm tr}({\bf 
\Phi }_{2}^{\dag}{\bf \Phi }_{2})\right \}^{2} 
 \nonumber \\
& &+\frac{1}{4}\lambda _{3}
{\rm tr}({\bf \Phi }_{1}^{\dag}
{\bf \Phi }_{1}){\rm tr}({\bf \Phi }_{2}^{\dag}
{\bf \Phi }_{2})  
+\frac{1}{16}\lambda _{4}\left 
\{{\rm tr}({\bf \Phi }_{1}^{\dag}{\bf 
\Phi }_{2})+{\rm tr}({\bf \Phi }_{2}^{\dag}{\bf 
\Phi }_{1})\right \}^{2} \nonumber \\
& &-\frac{1}{16}\lambda _{5}\left \{
{\rm tr}(\tau _{3}{\bf \Phi }_{2}^{\dag}{\bf \Phi }_{1})
-{\rm tr}(\tau _{3}{\bf \Phi }_{1}^{\dag}{\bf \Phi }_{2})
\right \}^{2}.
\end{eqnarray}
This shows clearly that, without the last term in Eq. (1),
the potential would possess  the global symmetry 
$SU(2)_{L}\times SU(2)_{R}$, 
under which ${\bf \Phi }_{i}$ undergoes the 
transformation 
${\bf \Phi }_{i}\rightarrow g_{L}{\bf \Phi }_{i}
g_{R}$ ($g_{L}\in SU(2)_{L}$, $g_{R}\in SU(2)_{R})$.
The  isospin symmetry 
$SU(2)_{V}={\rm diag}[SU(2)_{L}
\times SU(2)_{R}]$, that  would survive the 
spontaneous symmetry breaking, is also broken by the 
$\lambda _{5}$-interactions exclusively. 
We will come to this point later 
in connection with the screening theorem.

The particle content of the scalar sector may be 
seen  by putting
\begin{equation}
\Phi _{i}=\left (
\begin{array}{cc}
w_{i}^{+}   \\
\frac{1}{\sqrt{2}}(v_{i}+h_{i}+iz_{i})    
\end{array}
\right )
\end{equation}
and by shuffling Eq. (1).
Here $v_{i}$'s  ($i$=1, 2) are 
the vacuum expectation values.
The mass eigenstates are obtained by diagonalizing 
the quadratic terms in the neutral 
as well as charged sectors via 
\begin{equation}
\left (
\begin{array}{c}
h_{1} \\ h_{2}
\end{array}
\right )
=
\left (
\begin{array}{cc}
\cos \alpha & -\sin \alpha \\
\sin \alpha &  \cos \alpha \\
\end{array}
\right )
\left (
\begin{array}{c}
h \\ H
\end{array}
\right ),
\end{equation}

\begin{equation}
\left (
\begin{array}{c}
w_{1} \\ w_{2}
\end{array}
\right )
=
\left (
\begin{array}{cc}
\cos \beta & -\sin \beta  \\
\sin \beta &  \cos \beta   \\
\end{array}
\right )
\left (
\begin{array}{c}
w \\ G
\end{array}
\right ),
\end{equation}
\begin{equation}
\left (
\begin{array}{c}
z_{1} \\ z_{2}
\end{array}
\right )
=
\left (
\begin{array}{cc}
\cos \beta & -\sin \beta  \\
\sin \beta &  \cos \beta   \\
\end{array}
\right )
\left (
\begin{array}{c}
z \\ A
\end{array}
\right ).
\end{equation}
The mixing angle $\beta $ is simply given by 
$\tan \beta=v_{2}/v_{1}$.  As we can see easily,  
$H$ and $h$ are $CP$-even neutral
field, while $A$  a $CP$-odd neutral one.
The charged Higgs field is denoted by $G^{\pm}$, and 
the Nambu-Goldstone bosons are $w^{\pm}$ and $z$. 

The physical parameters of our theory are the masses 
$m_{H}$, $m_{h}$, $m_{G}$,  $m_{A}$ and the vacuum 
expectation value $v=\sqrt {v_{1}^{2}+v_{2}^{2}}$ 
together with the
mixing angles $\alpha$ and $\beta$. The quartic 
couplings in  (1) are expressed in terms of these 
physical parameters by

\begin{equation}
\lambda _{1}=\frac{1}{2v^{2}\cos ^{2}\beta}(m_{h}^{2}
\cos ^{2}\alpha +m_{H}^{2}\sin ^{2}\alpha ),
\end{equation}
\begin{equation}
\lambda _{2}=\frac{1}{2v^{2}\sin ^{2}\beta}(m_{h}^{2}
\sin ^{2}\alpha +m_{H}^{2}\cos ^{2}\alpha ),
\end{equation}
\begin{equation}
\lambda _{3}=\frac{\sin 2\alpha}{v^{2}\sin 2\beta}
(m_{h}^{2}-m_{H}^{2})+\frac{2m_{G}^{2}}{v^{2}},
\end{equation}
\begin{equation}
\lambda _{4}=-\frac{2m_{G}^{2}}{v^{2}},
\end{equation}
\begin{equation}
\lambda _{5}=\frac{2}{v^{2}}(m_{A}^{2}-m_{G}^{2}).
\end{equation}
\noindent
As wee see in  Eq. (11),
the deviation from the mass degeneracy 
$m_{G}^{2}=m_{A}^{2}$ 
between charged and {\it CP}-odd neutral Higgs 
bosons thus measures the custodial symmetry breaking. 

The quartic couplings are all assumed to be real and there
 is no source of $CP$-violation in the Higgs potential.
The $CP$-invariance enables us to set up several 
selection rules for triple and quartic Higgs couplings. 
It is worthwhile  mentioning    that
 we could  assign G-parity in the usual 
way as in hadron physics 
in connection with the isospin symmetry. The neutral
$h$ and  $H$ bosons have  even  G-parity, while 
$G$, $A$, $w$ and $z$  odd G-parity. 
This G-parity is also useful to 
set up selection rules for the Higgs-Goldstone 
 interactions for the case of $\lambda _{5}=0$.

\vskip2cm
\vfill\eject
\noindent 
{\large {\bf 3. The equivalence theorem at the 
loop level}}	
\vskip0.5cm 

The Higgs boson decay into a gauge boson pair is
 dominated preferentially by those into  longitudinally 
polarized ones if $m_{H}\gg M_{Z},M_{W}$. In such a case 
we may take an advantage  of the 
equivalence theorem [14-17]. This theorem states that 
the S-matrix elements  associated with 
longitudinal gauge bosons are approximated by those of
corresponding Nambu-Goldstone bosons;
\begin{eqnarray}
& &T(Z_{L}(p_{1}),\cdots  ,Z_{L}(p_{n}),
  W_{L}(q_{1}),\cdots  ,W_{L}(q_{m}))
                   \nonumber \\
&=&(C_{{\rm mod}}^{Z})^{n}(C_{{\rm mod}}^{W})^{m}
T(iz_(p_{1}), \cdots ,iz_(p_{n}), iw(q_{1}), \cdots 
 ,iw_(q_{m}))
+{\cal O}(M_{W}/\sqrt{s}).
                            \nonumber \\
\end{eqnarray}
Here $\sqrt{s}$ is the typical energy scale 
characterizing the scattering process 
and $C_{{\rm mod}}^{Z}$, and $C_{{\rm mod}}^{W}$
are  the so-called modification factor [17]
to be attached to each external gauge boson line of $Z$ 
and $W$'s, respectively. 

The equivalence theorem was first proved on the tree 
level [14, 15]. 
Since then, the validity of this theorem on the 
loop level has been examined by several authors [16-17]. 
The point is that the right-hand side of Eq. (12) is 
 unphysical (gauge-dependent) 
matrix elements and that we have to specify 
how to renormalize the external lines of the 
Nambu-Goldstone bosons. The modification factors 
$C_{{\rm mod}}^{Z}$ and $C_{{\rm mod}}^{W}$ 
are thus introduced to match the 
external line renormalization to the physical S-matrix 
 on the left-hand side. 
In a nutshell, if we  work in the on-shell 
renormalization scheme in the Landau gauge, these 
modification factors turn out at the one-loop level 
 to be [17]
\begin{equation}
C_{{\rm mod}}^{Z}=\sqrt{\frac{Z_{Z}}{Z_{z}}}
\sqrt{\frac{M_{Z}^{2}-\delta M_{Z}^{2}}{M_{Z}^{2}}},
\hskip1cm
C_{{\rm mod}}^{W}=\sqrt{\frac{Z_{W}}{Z_{w}}}
\sqrt{\frac{M_{W}^{2}-\delta M_{W}^{2}}{M_{W}^{2}}}.
\end{equation}

Here the wave function renormalization constants 
of the gauge bosons ($Z$ and $W^{\pm}$) and the 
Nambu-Goldstone  bosons ($z$ and $w^{\pm}$)are 
denoted by $Z_{Z}$, $Z_{W}$, $Z_{z}$ and $Z_{w}$, 
respectively.  The origin of Eq. (13) may be 
explained in the following very intuitive way. 
The factor $\sqrt{Z_{Z}/Z_{z}}$ and 
$\sqrt{Z_{W}/Z_{w}}$ compensates the difference 
in the external-line renormalization between  
Nambu-Goldstone  and the gauge bosons. The 
presence of $\delta M_{Z}^{2}$ and $\delta 
M_{W}^{2}$ in Eq. (13)  is due to the fact that 
gauge boson masses appearing in the longitudinal 
polarization vectors in the full theory are simply  
of kinematical origin irrelevant to renormalization,  
while they appear, in the 
Higgs-Goldstone systems, as coupling constants that 
are  subject to renormalization. 
Thus the  presence of $\delta M_{Z}^{2}$ and 
$\delta M_{W}^{2}$ in (13) may be undestood as making 
up this difference.  

Toussaint [20] once calculated $\delta M_{W}^{2}$ and 
$\delta M_{Z}^{2}$ at the one-loop level 
in the two Higgs doublet model, assuming 
that all the Higgs masses are much greater 
than the gauge bosons'. For completeness, we would like to 
include the effects due to the top quark mass 
since it is of  interest to see whether 
 the largeness of the top quark  mass might be just on the 
verge of  affecting the Higgs decays. 
After including  the top quark mass effects, Toussaint's  
 results on the vector boson mass renormalization are 
modified into
\begin{eqnarray}
\frac{\delta M_{Z}^{2}}{M_{Z}^{2}}&=&\frac{1}
{(4\pi )^{2}v^{2}}
\Biggl  \{
\frac{1}{2}(
m_{H}^{2}+m_{h}^{2}+m_{A}^{2}) \Biggr.
                 \nonumber \\
& &+\cos ^{2}(\alpha -\beta)\frac{m_{A}^{2}m_{H}^{2}}
{m_{H}^{2}-m_{A}^{2}}\ln\frac{m_{A}^{2}}{m_{H}^{2}}
+\sin ^{2}(\alpha -\beta)\frac{m_{A}^{2}m_{h}^{2}}
{m_{h}^{2}-m_{A}^{2}}\ln \frac{m_{A}^{2}}{m_{h}^{2}}
                    \nonumber \\
& &+\Biggl. 2N_{C}m_{t}^{2} (\frac{-2}
{D-4}-\gamma _{E}-\ln \frac{m_{t}^{2}}
{4\pi \mu ^{2}} ) \Biggr \},
\end{eqnarray}
\begin{eqnarray}
\frac{\delta M_{W}^{2}}{M_{W}^{2}}&=&
\frac{1}{(4\pi )^{2}v^{2}}  
\Biggl  \{ \frac{1}{2}(2m_{G}
^{2}+m_{H}^{2}+m_{h}^{2}+
m_{A}^{2})+\frac{m_{G}^{2}m_{A}^{2}}
{m_{A} ^{2}-m_{G}^{2}}\ln\frac{m_{G}^{2}}
{m_{A}^{2}}         \Biggr.
              \nonumber \\
& &+\cos ^{2}(\alpha -\beta)\frac{m_{G}^{2}m_{H}^{2}}
{m_{H}^{2}-m_{G}^{2}}\ln \frac{m_{G}^{2}}{m_{H}^{2}}
+\sin ^{2}(\alpha -\beta)\frac{m_{G}^{2}m_{h}^{2}}
{m_{h}^{2}-m_{G}^{2}}\ln \frac{m_{G}^{2}}{m_{h}^{2}}
                  \nonumber \\
& &+\Biggl. 2N_{C}m_{t}^{2} (\frac{-2}{D-4}+\frac{1}{2}
-\gamma _{E}-\ln \frac{m_{t}^{2}}{4\pi \mu ^{2}})
\Biggr \}.
\end{eqnarray}
Here $N_{C}$ is the number of colors ($N_{C}=3$), 
$\mu $  is the mass scale appearing in the $D$-dimensional 
regularization method, and here and hereafter 
we neglect the bottom quark mass effects. 

To compute the wave function renormalization constants 
$Z_{z}$ and $Z_{w}$ of the Nambu-Goldstone bosons and in 
particular the quark contributions to them,   we have 
summarized in Appendix A and Table 1 
the Yukawa couplings of the
 top and bottom quarks to various scalar particles. 
There are two models of the Yukawa coupling, 
so-called model I [21] and model II [22]. 
In model I, both top and bottom quarks receive their 
masses  from only one of the two Higgs doublets, say,
 $\Phi _{2}$. In the model II, on the other hand, 
the top quark mass comes from 
$\Phi _{2}$, and the bottom quark mass from $\Phi _{1}$.
As we see from the list of the couplings, 
there exsits little difference between these two models 
as far as the bottom quark mass is negligibly small. 
Setting $m_{b}\cong 0$, we will proceed hereafter without 
dicriminating the two models. 

The wave function renormalization constants 
$Z_{z}$, and $Z_{w}$ 
of the Nambu-Goldstone bosons are also easily 
extracted from the calculations in Ref. [20]
 together with those in Appendix C, namely,   
\begin{eqnarray}
Z_{z}&=&1-\frac{1}{(4\pi )^{2}v^{2}} 
\Biggl \{\frac{1}{2}(
m_{H}^{2}+m_{h}^{2}+
m_{A}^{2}) \Biggr.
                 \nonumber \\
& &+\cos ^{2}(\alpha -\beta)\frac{m_{A}^{2}m_{H}^{2}}
{m_{H}^{2}-m_{A}^{2}}\ln\frac{m_{A}^{2}}{m_{H}^{2}}
+\sin ^{2}(\alpha -\beta)\frac{m_{A}^{2}m_{h}^{2}}
{m_{h}^{2}-m_{A}^{2}}\ln \frac{m_{A}^{2}}{m_{h}^{2}}
                    \nonumber \\
& &-\Biggl. 2N_{C}m_{t}^{2}(\frac{2}{D-4}+\gamma _{E}
+\ln \frac{m_{t}^{2}}{4\pi \mu ^{2}}) \Biggr \},
\end{eqnarray}
\begin{eqnarray}
Z_{w}&=&1-\frac{1}{(4\pi )^{2}v^{2}} 
\Biggl \{
\frac{1}{2}(2m_{G}
^{2}+m_{H}^{2}+m_{h}^{2}+
m_{A}^{2})+\frac{m_{G}^{2}m_{A}^{2}}
{m_{A} ^{2}-m_{G}^{2}}\ln\frac{m_{G}^{2}}
{m_{A}^{2}} \Biggr.
       \nonumber \\
& &+\cos ^{2}(\alpha -\beta)\frac{m_{G}^{2}m_{H}^{2}}
{m_{H}^{2}-m_{G}^{2}}\ln\frac{m_{G}^{2}}{m_{H}^{2}}
+\sin ^{2}(\alpha -\beta)\frac{m_{G}^{2}m_{h}^{2}}
{m_{h}^{2}-m_{G}^{2}}\ln \frac{m_{G}^{2}}{m_{h}^{2}}
                  \nonumber \\
& &-\Biggl. 2N_{C}m_{t}^{2} (\frac{2}{D-4}+\gamma _{E}+
\ln \frac{m_{t}^{2}}{4\pi \mu ^{2}}-1 ) \Biggr \}.
\end{eqnarray}
The wave function renormalization 
constants of the $W$ and $Z$
  bosons, on the other hand,  are given by 
$Z_{Z}=1+{\cal O}(M_{Z}^{2}/v^{2})$, and 
$Z_{W}=1+{\cal O}(M_{W}^{2}/v^{2})$, and may be set equal 
to unity in our approximation scheme.  In summary, 
the modification factors in (13)  turn out to be
\begin{equation}
C_{{\rm mod}}^{Z}\cong 1, 
\hskip1cm
C_{{\rm mod}}^{W}\cong 1+\frac{1}
{(4\pi )^{2}v^{2}}\cdot 
\frac{N_{C}m_{t}^{2}}{2}.
\end{equation}

\vskip1cm
\noindent 
{\large {\bf 4. The Higgs-Goldstone system
}}
\vskip0.5cm

Being equipped with the machinery of the 
equivalence theorem,  we are now 
interested in the radiative corrections to the 
processes 
$H\rightarrow zz$ and $H\rightarrow w^{+}w^{-}$.
Since we  assume  that all the Higgs boson 
masses are much greater than $M_{Z}$ and $M_{W}$, 
internal loops are also dominated by Higgs and 
Nambu-Goldstone bosons (plus top quark). 

Before starting to calculate the radiative corrections, 
we have to go somewhat in detail on the structure of the 
counterterms, since there are subtleties with regard to the 
field mixing. 
The interaction Lagrangian relevant to $Hzz$ and 
$Hw^{+}w^{-}$ vertices is
extracted from the Higgs potential (1)  
\begin{equation}
{\cal L}_{Hzz}=\frac{m_{H}^{2}}{2v}\sin (\alpha 
-\beta)Hzz, 
\end{equation}
\begin{equation}
{\cal L}_{Hww}=\frac{m_{H}^{2}}{v}\sin  (\alpha 
-\beta)Hw^{+}w^{-}.
\end{equation}
The counterterms, 
$\delta {\cal L}_{Hzz}$, and
$\delta {\cal L}_{Hww}$,  
 consist of two parts, {\it i.e.},
\begin{equation}
\delta {\cal L}_{Hzz}=\delta {\cal L}_{Hzz}^{(1)}+ 
\delta {\cal L}_{Hzz}^{(2)}, 
\hskip1cm
\delta {\cal L}_{Hww}=\delta {\cal L}_{Hww}^{(1)}+  
\delta {\cal L}_{Hww}^{(2)}.
\end{equation}
The first one 
is obtained simply by replacing the parameters in (19)
and (20)  as 
$m_{H}^{2}\rightarrow m_{H}^{2}-\delta m_{H}^{2}$, 
$\alpha\rightarrow  \alpha -\delta \alpha$, 
$\beta \rightarrow \beta -\delta \beta$, 
$v\rightarrow v-\delta v$. We thus obtain
\begin{eqnarray}
\delta {\cal L}_{Hzz}^{(1)}&=&
\left \{-\frac{\delta m_{H}^{2}}
{m_{H}^{2}}+\frac{\delta v}{v}\right \}
\frac{m_{H}^{2}}{2v}
\sin (\alpha -\beta )Hzz 
                          \nonumber \\
& &-\frac{m_{H}^{2}}{2v}(\delta \alpha -\delta \beta)
\cos (\alpha -\beta)Hzz,
\end{eqnarray}
\begin{eqnarray}
\delta {\cal L}_{Hww}^{(1)}&=&\left 
\{-\frac{\delta m_{H}^{2}}
{m_{H}^{2}}+\frac{\delta v}{v}\right \}\frac{m_{H}^{2}}{v}
\sin (\alpha -\beta)Hw^{+}w^{-}
                   \nonumber \\
& &-\frac{m_{H}^{2}}{v}
(\delta \alpha -\delta \beta)\cos (\alpha -\beta )
Hw^{+}w^{-}.          
\end{eqnarray}

The second one is rather complicated due to 
field mixing effects between 
$H\leftrightarrow h$, 
$z\leftrightarrow A$, and $w^{\pm}\leftrightarrow G^{\pm}$ 
pairs. These mixing effects are described by  $2\times 2$ 
wave function renormalization matrices together with 
the mixing angle renormalization, $\delta \alpha $ and
$\delta \beta $. The renormalization 
is fullfilled by the replacement 
\begin{equation}
\left (
\begin{array}{c}
h \\
H \\
\end{array}
\right )
\rightarrow
\left (
\begin{array}{cc}
\sqrt{Z_{h}} & \sqrt{Z_{hH}} \\
\sqrt{Z_{Hh}} & \sqrt{Z_{H}} \\
\end{array}
\right )
\left (
\begin{array}{cc}
\cos \delta \alpha & -\sin \delta \alpha  \\
\sin \delta \alpha &  \cos \delta \alpha  \\
\end{array}
\right )
\left (
\begin{array}{c}
h \\
H \\
\end{array}
\right )
,
\end{equation}
\begin{equation}
\left (
\begin{array}{c}
z \\
A \\
\end{array}
\right )
\rightarrow
\left (
\begin{array}{cc}
\sqrt{Z_{z}} & \sqrt{Z_{zA}} \\
\sqrt{Z_{Az}} & \sqrt{Z_{A}} \\
\end{array}
\right )
\left (
\begin{array}{cc}
\cos \delta \beta & -\sin \delta \beta  \\
\sin \delta \beta &  \cos \delta \beta  \\
\end{array}
\right )
\left (
\begin{array}{c}
z \\
A \\
\end{array}
\right )
,
\end{equation}
\begin{equation}
\left (
\begin{array}{c}
w \\
G \\
\end{array}
\right )
\rightarrow
\left (
\begin{array}{cc}
\sqrt{Z_{w}} & \sqrt{Z_{wG}} \\
\sqrt{Z_{Gw}} & \sqrt{Z_{G}} \\
\end{array}
\right )
\left (
\begin{array}{cc}
\cos \delta \beta & -\sin \delta \beta  \\
\sin \delta \beta &  \cos \delta \beta  \\
\end{array}
\right )
\left (
\begin{array}{c}
w \\
G \\
\end{array}
\right )
.
\end{equation}

Because of these mixings, 
the counterterms for the $Hzz$ vertex  
are also produced from the $hzz$ and $HAz$ terms in the
 Lagrangian, {\it i.e.},  
\begin{equation}
{\cal L}_{hzz}=-\frac{m_{h}^{2}}{2v}\cos (\alpha 
-\beta)hzz, 
\end{equation}
\begin{equation}
{\cal L}_{HAz}=-\frac{m_{H}^{2}-m_{A}^{2}}{v}\cos
 (\alpha -\beta)HAz. 
\end{equation}
Similarly, the interactions coming from the potential (1)
\begin{equation}
{\cal L}_{hww}=-\frac{m_{h}^{2}}{v}\cos (\alpha 
-\beta)hw^{+}w^{-}, 
\end{equation}
\begin{equation}
{\cal L}_{HGw}=-\frac{m_{H}^{2}-m_{G}^{2}}{v}\cos
 (\alpha -\beta)(Hw^{+}G^{-}+Hw^{-}G^{+}), 
\end{equation}
provide some of the counterterms for $Hw^{+}w^{-}$ 
interaction. 
We will restrict our consideration to the lowest-order 
loop-corrections and hence keep only linear terms in 
$\delta \alpha $ and $\delta \beta $.
It is in fact straightforward by the replacement (24), 
(25) and (26) 
in the interactions (27)-(30), to reach the following set 
of counterterms which are to be added to (22) and (23);
\begin{eqnarray}
\delta {\cal L}_{Hzz}^{(2)}&=&\left \{(\sqrt{Z_{H}}-1)+
(Z_{z}-1)\right \}\frac{m_{H}^{2}}{2v}\sin (\alpha -\beta)Hzz
                      \nonumber \\
& &-(\sqrt{Z_{hH}}-\delta \alpha)\frac{m_{h}^{2}}{2v}\cos 
(\alpha -\beta )Hzz        \nonumber \\
& &-(\sqrt{Z_{Az}}+\delta \beta)\frac{m_{H}^{2}-m_{A}^{2}}
{v}\cos (\alpha -\beta )Hzz,
                            \nonumber \\
\end{eqnarray}
\begin{eqnarray}
\delta {\cal L}_{Hww}^{(2)}&=&
\left \{(\sqrt{Z_{H}}-1)+(Z_{w}-1)\right \}\frac{m_{H}^{2}}
{v}\sin (\alpha -\beta )Hw^{+}w^{-}
                               \nonumber \\
& &-(\sqrt{Z_{hH}}-\delta \alpha )\frac{m_{h}^{2}}{v}
\cos  (\alpha -\beta )Hw^{+}w^{-}
                                \nonumber \\
& &-2(\sqrt{Z_{Gw}}+\delta \beta)\frac{m_{H}^{2}-
m_{G}^{2}}{v}\cos (\alpha -\beta )Hw^{+}w^{-}.
                      \nonumber \\
\end{eqnarray}

It is by now clear what kinds of two-point 
functions are to be computed in order to 
complete our counterterms.
Hereafter we will use the notations 
$\Pi _{ij}(p^{2})$ for the two point functions 
corresponding to Fig. 1, 
where indices $i$ and $j$ denote either one of 
the scalar field, $H$, $h$, 
$z$, $w$, $A$ or $G$.  
Internal particles in Fig. 1(a) and 1(b) are given 
in Table 2. 
Calculations of some of the two-point functions are 
sketched in Appendices C and D. 
The basic quantities in  the 
counterterms are summarized as follows:
\begin{eqnarray}
\delta m_{H}^{2}&=&{\rm Re}\left (
\Pi _{HH}(m_{H}^{2})\right ),
\hskip1cm 
Z_{H}=1+{\rm Re }\left (\Pi '_{HH}(m_{H}^{2})\right ), 
                               \nonumber \\
Z_{z}&=&1+{\rm Re}\left (\Pi '_{zz}(0)\right ), 
\hskip1cm
Z_{w}=1+{\rm Re }\left (\Pi '_{ww}(0)\right ).
\end{eqnarray}
Here the renormalization is  done on the mass shell, 
namely, at $p^{2}=m_{H}^{2}$ for 
$\Pi _{HH}$ and $p^{2}=0$ for 
$\Pi _{zz}$ and $\Pi _{ww}$.
The divergences in the mixing of the 
two-point function $\Pi _{hH}$ 
are taken care of by the counterterms $\delta \alpha$, 
$\sqrt{Z_{hH}}$ and $\sqrt{Z_{Hh}}$. 
In general, $Z_{hH}$ and $Z_{Hh}$ are not bound to be 
identical. We will, however, impose $Z_{hH}=Z_{Hh}$ just for 
simplicity. If the subtraction is 
carried out at $p^{2}=m_{H}^{2}$, they are determined by
\begin{equation}
\delta \alpha =\frac{1}{m_{h}^{2}-m_{H}^{2}}
{\rm Re}\left (\Pi _{hH}(m_{H}^{2})\right )
+\frac{1}{2}{\rm Re}\left (\Pi '_{hH}(m_{H}^{2})\right ),
\end{equation}
\begin{equation}
\sqrt{Z_{hH}}=\sqrt{Z_{Hh}}=\frac{1}{2}{\rm Re}\left (
\Pi '_{hH}(m_{H}^{2})\right ).
\end{equation}
In the same way, the remainimg counterterms  are 
determined by 
\begin{equation}
\delta \beta =-\frac{1}{m_{A}^{2}}{\rm Re}\left (
\Pi _{zA}(0)\right )-\frac{1}{2}{\rm Re }\left (
\Pi '_{zA}(0)\right )=-\frac{1}{2}{\rm Re}\left (
\Pi '_{zA}(0)\right ),
\end{equation}
\begin{equation}
\sqrt{Z_{zA}}=\sqrt{Z_{Az}}=\frac{1}{2}{\rm Re }
\left (\Pi '_{zA}(0)\right ),
\hskip1cm
\sqrt{Z_{wG}}=\sqrt{Z_{wG}}=\frac{1}{2}{\rm Re}
\left (\Pi '_{wG}(0)\right ),
\end{equation}
In Eq. (36), we have used the fact that
 $\Pi _{zA}(0)$ vanishes
 as a consequence of the  Nambu-Goldstone theorem.
The renormalization of $\beta $ could be done 
alternatively by using $\Pi _{wG}(0)$. The difference 
between $\Pi _{zA}(0)$ and $\Pi _{wG}(0)$ is of course 
finite and infinities may be eliminated no matter 
whichever choice   we would take.
The use of $\Pi _{zA}$  instead of $\Pi _{wG}$ 
to determine $\delta \beta$
is just our convention. 

We should add a few words as to the counterterm $\delta v$ 
to the vacuuum expectation value. In the on-shell 
renormalization scheme [23] in which  $M_{Z}$, $M_{W}$, 
 and the fine structure constant 
($e^{2}/4\pi$) are physical input parameters, 
$\delta v$ is determined by 
\begin{equation}
\frac{\delta v}{v}=
\frac{1}{2}\left(
\frac{\delta M_{W}^{2}}{M_{W}^{2}}-\frac{\delta 
M_{Z}^{2}}{M_{Z}^{2}}+\frac{\delta M_{Z}^{2}-\delta 
M_{W}^{2}}{M_{Z}^{2}-M_{W}^{2}}-\frac{\delta e^{2}}
{e^{2}}\right).
\end{equation}
There is an alternative way which is referred to as 
modified on-shell scheme [24] and uses the muon decay 
constant instead of $M_{W}$ as an input. 
 This scheme was used in our previous work [5]. 
We have confirmed that the difference 
between the two prescriptions is small numerically. 
Just for definiteness, we use (38)  in our 
numerical calculations.

\vskip1cm
\vfill\eject
\noindent 
{\large {\bf 5. Loop corrections to the vertices}}
\vskip0.5cm

We are now in a position to carry out 
 the loop calculations of  
 the $Hzz$ vertex together with $Hw^{+}w^{-}$'s. 
The Feynman diagrams contributing to the renormalization 
of the $Hzz$ vertex are depicted in Fig. 2.
They are classified into scalar part and top quark part:
\begin{equation}
\Gamma _{Hzz}(p^{2})=\Gamma _{Hzz}^{({\rm scalar})}
(p^{2})+N_{C}\Gamma _{Hzz}^{({\rm quark})}(p^{2}).
\end{equation}
The scalar loop contributions are divided further 
into three terms
\begin{equation}
\Gamma _{Hzz}^{({\rm scalar})}(p^{2})=\Gamma _{Hzz}^{(1)}
+\Gamma _{Hzz}^{(2)}+\Gamma _{Hzz}^{(3)}
\end{equation}
according to the types of Feynman diagrams in Fig. 2.
Internal particles in Fig. 2 are listed in Table 3.

The calculation is tedious but straightforward 
and we just record the results. The Feynman diagrams 
corresponding to Fig. 2(a)
are summed up to:
\begin{eqnarray}
\Gamma _{Hzz}^{(1)}&=&
-\frac{m_{H}^{6}}{v^{3}}\sin ^{3}(
\alpha -\beta)g(p^{2}, 0, 0, m_{H}^{2}) 
                                    \nonumber \\
& &-\frac{m_{h}^{4}m_{H}^{2}}{v^{3}}\sin 
(\alpha -\beta)\cos ^{2}(\alpha -\beta) g(p^{2}, 0, 0,
 m_{h}^{2})    \nonumber \\
& &+\frac{1}{v^{3}}\Biggl \{
m_{H}^{2}(\frac{\cos \alpha \cos ^{2}\beta
}{\sin \beta}-\frac{\sin \alpha \sin ^{2}\beta
}{\cos \beta})   -2m_{A}^{2}\sin (\alpha -\beta)
\Biggr \}
    \nonumber  \\
& &\times \Biggl \{(m_{h}^{2}-m_{A}^{2})^{2}\sin ^{2}
(\alpha -\beta)g(p^{2}, m_{A}^{2},  m_{A}^{2},  m_{h}^{2})
 \Biggr.   \nonumber \\
& &+\Biggl. (m_{H}^{2}-
m_{A}^{2})^{2}\cos ^{2}(\alpha -\beta)
g(p^{2},  m_{A}^{2},  m_{A}^{2},  m_{H}^{2})
\Biggr \}      \nonumber \\
& &+\frac{2m_{h}^{2}(m_{H}^{2}-m_{A}^{2})(m_{h}^
{2}-m_{A}^{2})}{v^{3}}\sin (\alpha -\beta)
\cos ^{2}(\alpha -\beta)
g(p^{2},  0,  m_{A}^{2},  m_{h}^{2})
   \nonumber \\
& &-\frac{2m_{H}^{2}(m_{H}^{2}-m_{A}^{2})^{2}
}{v^{3}}\sin (\alpha -\beta)
\cos ^{2}(\alpha -\beta)
g(p^{2},  0,  m_{A}^{2},  m_{H}^{2})
   \nonumber \\
& &+\frac{3m_{H}^{6}}{v^{3}}\sin ^{2}(\alpha -
\beta)(\frac{\cos ^{3}\alpha }{\sin \beta}-
\frac{\sin ^{3}\alpha}{\cos \beta})g(p^{2}, 
  m_{H}^{2},  m_{H}^{2},  0)
       \nonumber \\
& &+\frac{3m_{H}^{2}(m_{H}^{2}-m_{A}^{2})
^{2}}{v^{3}}\cos ^{2}(\alpha -
\beta)(\frac{\cos ^{3}\alpha }{\sin \beta}-
\frac{\sin ^{3}\alpha}{\cos \beta})g(p^{2}, 
  m_{H}^{2},  m_{H}^{2},  m_{A}^{2})
       \nonumber \\
& &-\frac{2m_{h}^{2}m_{H}^{2}(2m_{H}^
{2}+m_{h}^{2})}{v^{3}}
\sin (\alpha -\beta)\cos ^{2}(\alpha -\beta)
\frac{\sin 2\alpha}{\sin 2\beta}
g(p^{2},  m_{H}^{2},  m_{h}^{2},  0)
      \nonumber \\
& &+\frac{2(m_{h}^{2}-m_{A}^{2})(m_{H}^{2}
-m_{A}^{2})(2m_{H}^{2}+m_{h}^{2})}{v^{3}}
\sin (\alpha -\beta)\cos ^{2}(\alpha -\beta)
\frac{\sin 2\alpha}{\sin 2\beta}
       \nonumber \\
& &\times g(p^{2},  m_{H}^{2},  m_{h}^{2},  m_{A}^{2})
      \nonumber \\
& &+\frac{m_{h}^{4}(m_{H}^{2}+2m_{h}^{2})}{v^{3}}
\sin (\alpha -\beta)\cos ^{2}(\alpha -\beta)
\frac{\sin 2\alpha}{\sin 2\beta}
g(p^{2},  m_{h}^{2},  m_{h}^{2},  0)
                           \nonumber \\
& &+\frac{(m_{h}^{2}-m_{A}^{2})^{2}
(m_{H}^{2}+2m_{h}^{2})}{v^{3}}
\sin ^{3}(\alpha -\beta) 
\frac{\sin 2\alpha}{\sin 2\beta}
g(p^{2},  m_{h}^{2},  m_{h}^{2},  m_{A}^{2}).
                           \nonumber \\
\end{eqnarray}
Each term of the above expression is easily 
identified with the corresponding Feynman diagrams 
just by looking at the function $g(p^{2}, m_{1}^{2},
m_{2}^{2}, m_{3}^{2})$ defined in Appendix B.
Those diagrams of the type of Fig. 2(b) are, on the 
other hand, expressed by the integral $f_{2}(p^{2}, 
m_{1}^{2}, m_{2}^{2})$ which is also defined in 
Appendix B. They turn out 
to be:
\begin{eqnarray}
\Gamma _{Hzz}^{(2)}&=&
\frac{5m_{H}^{2}}{2v^{3}}
\Biggl \{m_{h}^{2}\cos ^{2}(\alpha 
-\beta)+m_{H}^{2}\sin ^{2}(\alpha -\beta)\Biggr \}
\sin (\alpha -\beta)f_{2}(p^{2},  0,  0)
         \nonumber \\
& &-\frac{1}{2v^{3}}\Biggr \{
m_{H}^{2}(\frac{\cos \alpha\cos 
^{2}\beta}{\sin \beta}-\frac{\sin \alpha \sin ^{2}
\beta}{\cos \beta})-2m_{A}^{2}\sin (\alpha -\beta)
\Biggr \}        \nonumber \\
& &\times \Biggl \{
3m_{h}^{2}\sin ^{2}(\alpha -\beta)+3m_{H}^{2}
\cos ^{2}(\alpha -\beta)\Biggr.
           \nonumber \\
& &+\Biggl. \frac{\sin 2\alpha}{\sin 2\beta}
(m_{h}^{2}-m_{H}^{2})\Biggr \}f_{2}(p^{2}, 
m_{A}^{2},  m_{A}^{2})  \nonumber \\
& &-\frac{2}{v^{3}}\Biggl \{m_{H}^{2}(\frac{\cos \alpha\cos 
^{2}\beta}{\sin \beta}-\frac{\sin \alpha \sin ^{2}
\beta}{\cos \beta})-2m_{G}^{2}\sin (\alpha -\beta)
\Biggr \}        \nonumber \\
& &\times \Biggl \{
\frac{1}{2}m_{h}^{2}\sin ^{2}(\alpha -\beta)+\frac
{1}{2}m_{H}^{2}\cos ^{2}(\alpha -\beta) \Biggr.
           \nonumber \\
& &+\Biggl. \frac{1}{2}\frac{\sin 2\alpha}{\sin 2\beta}
(m_{h}^{2}-m_{H}^{2})+m_{G}^{2}\Biggr \}f_{2}(p^{2}, 
m_{G}^{2},  m_{G}^{2})   \nonumber \\
& &-\frac{2}{v^{3}}(m_{h}^{2}-m_{H}^{2})
(m_{H}^{2}-m_{G}^{2})\sin (\alpha -\beta)
\cos ^{2}(\alpha -\beta)f_{2}(p^{2},  m_{G}^{2},
  0)
             \nonumber \\
& &-\frac{3}{v^{3}}(m_{h}^{2}-m_{H}^{2})
(m_{H}^{2}-m_{A}^{2})\sin (\alpha -\beta)
\cos ^{2}(\alpha -\beta)f_{2}(p^{2},  m_{A} ^{2},
  0)
             \nonumber \\
& &-\frac{3m_{H}^{2}}{2v^{3}}
\Biggl \{
\left ((m_{h}^{2}-m_{H}^{2})\frac{\sin 2\alpha}{\sin 2\beta}
+2m_{A}^{2}\right )\cos ^{2}(\alpha -\beta)+
 m_{H}^{2}\Biggr \}
               \nonumber \\
& &\times \left (\frac{\cos ^{3}\alpha}
{\sin \beta}-\frac{\sin ^{3}\alpha}{\cos \beta}\right )
f_{2}(p^{2},  m_{H}^{2},  m_{H}^{2})
             \nonumber \\
& &-\frac{1}{v^{3}}\Biggl \{(m_{h}^{2}-m_{H}^{2})\frac{
\sin 2\alpha}{\sin 2\beta}+2m_{A}^{2}\Biggr  \}(m_{h}^{2}
+2m_{H}^{2})
              \nonumber \\
& &\times \sin (\alpha -\beta)\cos ^{2}(\alpha -
\beta)\frac{\sin 2\alpha }{\sin 2\beta}
f_{2}(p^{2},  m_{H}^{2},   m_{h}^{2})
        \nonumber \\
& &-\frac{1}{2v^{3}}\Biggl \{\left ((m_{h}^{2}-
m_{H}^{2})\frac{
\sin 2\alpha}{\sin 2\beta}+2m_{A}^{2}\right )\sin ^{2}
(\alpha -\beta)+m_{h}^{2}
\Biggr \}            \nonumber \\
& &\times (m_{H}^{2}+2m_{h}^{2})\sin (\alpha -\beta )
\frac{\sin 2\alpha }{\sin 2\beta}f_{2}(p^{2},  
m_{h}^{2},   m_{h}^{2}).
              \nonumber \\
\end{eqnarray}
Finally we come to the sum of Feynman diagrams 
of the type Fig. 2(c):
\begin{eqnarray}
\Gamma _{Hzz}^{(3)}&=&
-\frac{2m_{h}^{2}}{v^{3}}\Biggl  \{(m_{h}^{2}-
m_{H}^{2})\frac{\sin 2\alpha}{\sin 2\beta}
+2m_{A}^{2}\Biggr  \}
                       \nonumber \\
& &\times \sin (\alpha -\beta )
\cos ^{2}(\alpha -\beta )f_{2}(0, m_{h}^{2}, 0)
              \nonumber \\
& &+\frac{2m_{H}^{2}}{v^{3}}\Biggl \{(m_{h}^{2}
\frac{\sin 2\alpha }{\sin 2\beta}+2m_{A}^{2})
\sin (\alpha -\beta)\cos ^{2}(\alpha -\beta)
\Biggr.            \nonumber  \\
& & +\Biggl. m_{H}^{2}
(\frac{\sin ^{3}\alpha }{\cos \beta}
-\frac{\cos ^{3}\alpha}{\sin \beta})\sin ^{2}(\alpha
-\beta)\Biggr \}f_{2}(0, m_{H}^{2}, 0)
      \nonumber \\
& &-\frac{2(m_{h}^{2}-m_{A}^{2})}{v^{3}}
\Biggl \{\frac{\sin 2\alpha}{\sin 2\beta }(m_{h}^{2}
\sin ^{2}(\alpha -\beta )
+m_{H}^{2}\cos ^{2}(\alpha -\beta)
)\Biggr. 
\nonumber \\
& &-\Biggl. m_{A}^{2}\cos (2\alpha 
-2\beta)\Biggr \}\sin (\alpha -\beta)f_{2}(0, 
m_{A}^{2}, m_{h}^{2})
           \nonumber \\
& & -\frac{m_{H}^{2}-m_{A}^{2}}{v^{3}}\Biggl \{
(m_{h}^{2}\frac{\sin 2\alpha}{\sin 2\beta}+
2m_{A}^{2})\sin (2\alpha -2\beta)\Biggr. 
      \nonumber \\
& &+\Biggl. 2m_{H}^{2}(\frac{\cos ^{3}\alpha}{\sin \beta}
-\frac{\sin ^{3}\alpha}{\cos \beta})
\cos (\alpha -\beta)\Biggr \}
\cos (\alpha -\beta)f_{2}(0, m_{A}^{2}, m_{H}^{2}).
     \nonumber \\
\end{eqnarray}
The top quark contributions are obtained by using the 
Yukawa couplings listed  in Table 1:
\begin{equation}
\Gamma _{Hzz}^{({\rm quark})}=
\frac{4m_{t}^{4}}{v^{3}}\frac{\cos \alpha }{\sin \beta}
\left \{2f_{2}(p^{2},  m_{t}^{2},  
m_{t}^{2})-p^{2}g(p^{2},  m_{t}^{2},  
m_{t}^{2}, m_{t}^{2})\right \}.
\end{equation}

The loop corrections to the $Hw^{+}w^{-}$ vertex go 
in the same way and we again separate the 
contributions into scalar and top quark parts,
\begin{equation}
\Gamma _{Hww}(p^{2})=\Gamma _{Hww}^{({\rm scalar})}(p^{2})
+N_{C}\Gamma _{Hww}^{({\rm quark})}(p^{2}).
\end{equation}
The scalar contributions 
$\Gamma _{Hww}^{({\rm scalar})}(p^{2})$
were computed in our previous paper  (see Eq. (31)
 in Ref. [5]), and we need not reproduce them here. 
The top and bottom quark part is given, if we set
$m_{b}=0$,  by 
\begin{eqnarray}
\Gamma _{Hww}^{({\rm quark})}&=&
\frac{4m_{t}^{4}}{v^{3}}\frac{\cos \alpha}{\sin \beta}
\left \{f_{2}(p^{2},  m_{t}^{2}, m_{t}^{2})+f_{2}(0, 0,  
m_{t}^{2})\right \}
                  \nonumber \\
& &-\frac{4m_{t}^{6}}{v^{3}}\frac{\cos \alpha}{\sin \beta}
g(p^{2},  m_{t}^{2}, m_{t}^{2}, 0).
                    \nonumber \\
\end{eqnarray}

By adding the counterterm contributions we 
end up with the  decay width formulae,
\begin{equation}
\Gamma (H\rightarrow Z_{L}Z_{L})=
\frac{1}{32\pi}\frac{1}{m_{H}}\sqrt{1-\frac{
4M_{Z}^{2}}{m_{H}^{2}}}\vert {\cal M}_{Hzz}(p^{2}=m_{H}^{2}
) \vert ^{2} \vert C_{{\rm mod}}^{Z}\vert ^{4},
\end{equation}
\begin{equation}
\Gamma (H\rightarrow W_{L}^{+}W_{L}^{-})=
\frac{1}{16\pi}\frac{1}{m_{H}}\sqrt{1-\frac{
4M_{W}^{2}}{m_{H}^{2}}}\vert {\cal M}_{Hww}(p^{2}=m_{H}^{2}
) \vert ^{2}\vert C_{{\rm mod}}^{W}\vert ^{4},
\end{equation}
where the invariant amplitudes are given by
\begin{eqnarray}
{\cal M}_{Hzz}(p^{2})&=& \Gamma _{Hzz}(p^{2})+
 \frac{1}{v}\cos (\alpha -\beta)
{\rm Re}\left (\Pi _{hH}(m_{H}^{2})\right )
                              \nonumber \\
& &-\frac{1}{v}\sin (\alpha -\beta)
{\rm Re}\left (\Pi _{HH}(m_{H}^{2})\right )    
                        \nonumber \\
& &+\left \{\frac{\delta v}{v} +\frac{1}{2}{\rm Re}
\left (\Pi _{HH}'(m_{H}^{2})\right )
+Z_{z}\right \}\frac{m_{H}^{2}}{v}\sin (\alpha 
-\beta)
                      \nonumber \\
& &-{\rm Re}\left (\Pi '_{zA}(0)+\Pi '_{hH}(m_{H}^{2})
\right )
\frac{m_{H}^{2}}{2v}\cos (\alpha -\beta),
\end{eqnarray}
\begin{eqnarray}
{\cal M}_{Hww}(p^{2})&=& \Gamma _{Hww}(p^{2})+ 
\frac{1}{v}\cos (\alpha -\beta)
{\rm Re}\left (\Pi _{hH}(m_{H}^{2})\right )
                        \nonumber \\
& &-\frac{1}{v}\sin (\alpha -\beta)
{\rm Re}\left (\Pi _{HH}(m_{H}^{2}) \right )
                   \nonumber \\
& &+\left \{\frac{\delta v}{v} +\frac{1}{2}
{\rm Re}\left (\Pi _{HH}'(m_{H}^{2})\right )
+Z_{w}\right \}\frac{m_{H}^{2}}{v}\sin (\alpha 
-\beta)
                          \nonumber \\
& &-{\rm Re}\left (\Pi '_{hH}(m_{H}^{2})+
\Pi '_{wG}(0)\right )
\frac{m_{H}^{2}}{2v}\cos (\alpha -\beta)
                          \nonumber \\
& &+{\rm Re}\left (\Pi '_{wG}(0)-\Pi '_{zA}(0)
\right )\frac{2m_{G}^{2}-m_{H}^{2}}
{2v}\cos (\alpha -\beta).
\end{eqnarray}
These amplitudes are necessarily  finite and the 
finiteness is a non-trivial check of the calculations.

\vskip1cm
\noindent 
{\large {\bf 6. Numerical analyses of the 
decay widths}}
\vskip0.5cm

Let us now analyze the decay width formulae 
 numerically and look at their heavy Higgs boson 
mass-dependences. In doing so we have to select 
reasonable  numbers for the set of parameters,   
$m_{H}$, $m_{G}$, $m_{A}$, $m_{h}$, $\alpha $ 
and $\beta $.  There are four kinds of experimental 
information that we have to bear in our mind:
(1) the measurement of the $\rho$-parameter [20, 25], 
(2) the neutral meson mixings (
$K^{0}-{\bar K}^{0}$, $B^{0}-{\bar B}^{0}$, and
$D^{0}-{\bar D}^{0}$) [26],
(3) the recent measurement 
of the decays such as 
  $B\rightarrow K^{*}(892)\gamma $ [27],
(4) the ratios, 
$R_{b}=\Gamma (Z\rightarrow b{\bar b})/\Gamma 
(Z\rightarrow {\rm hadrons})$, and the 
$R_{c}$ counterpart [28].

It has been  known rather well that the constraints 
on the deviation of the $\rho$-parameter from unity 
prohibits $m_{G}$  to be much larger than or much 
smaller than either of neutral Higgs boson masses.  
These constraints on the Higgs masses, however, 
depend on the mixing angle, $\alpha $ and $\beta$. 
The simultaneous analysis on the masses and mixing  
angles would become much involved. For the purpose  
of getting an insight into a global picture of the  
$m_{G}$- and $m_{A}$-dependences of the decay widths, 
we may well vary the masses , $m_{G}$ and $m_{A}$, 
a little beyond the $\rho$-parameter constraint.
The data on neutral meson mixings and 
$b\rightarrow s\gamma$ decay rate both rule out small
 value of  $m_{G}$ and small $\tan \beta$. 
Grant [29] has confronted the two-doublet model with 
these experimental data simultaneously and made an 
overall analyses. The constraints he derived are 
$m_{G}>150-200 {\rm GeV}$ and $\tan \beta >0.7$. 
We will give these values due considerations  in 
the following numerical analyses. 
 The so-called $R_{b}-R_{c}$ crisis [28] , which might
 jeopardise  the standard model, is our  recent central
 concern. It seems, however, yet premature to draw 
definite  conclusions from those data or implications 
to the two-doublet model.  We will henceforth wait a little
while for what comes next.

Besides these sets of experimental information, 
we have theoretical contraints on the masses in the
 two-doublet model. One of them is the triviality bound 
 [30] and another is the tree unitarity bound [31]. 
The latter constraints lead to  $m_{H}<500$  GeV,
 $m_{G}<870$ GeV, $m_{h}< 710$ GeV and $m_{A}<1200$ 
GeV.   The triviality bounds also
 provide roughly similar results. We will take  these 
into our consideration as a guide of our parameter 
choices.  We will set $m_{H}=300$  GeV, and $m_{h}=400$ 
GeV throughout our numerical calculations.

First of all, we look at the 
$m_{G}$- and $m_{A}$-dependences 
of $\Gamma (H\rightarrow Z_{L}Z_{L})$ and   
 $\Gamma (H\rightarrow W^{+}_{L}W^{-}_{L})$ 
for the following 
two cases,
\begin{description}
\item{Case I:}
$\tan \beta =2$ and $\sin ^{2}(\alpha -\beta)=1$,
\item{Case II:}
$\tan \beta =10$ and $\sin ^{2}(\alpha -\beta)=
0.5 $.
\end{description}
The decay widths 
$\Gamma(H\rightarrow Z_{L}Z_{L})$
and $\Gamma(H\rightarrow W^{+}_{L}W^{-}_{L})$ 
as a function of $m_{G}$ depicted in Figs.  3 and 4 
correspond to the above two cases.
We vary $m_{G}$ from  300  GeV to 1000 GeV, while 
$m_{A}$ is  set equal to 400, 600 and 800 GeV. 

To get a rough  idea of numerical values  let us recall 
that these decay rates in the minimal standard model 
(MSM) with a single Higgs doublet
 give at the tree level 
\begin{eqnarray}
\Gamma ^{{\rm MSM}}(H\rightarrow Z_{L}Z_{L}) 
\vert _{{\rm tree}}
&=&\frac{1}{32\pi }\frac{m_{H}^{3}}{v^{2}}\sqrt{
1-\frac{4M_{Z}^{2}}{m_{H}^{2}}},
         \nonumber \\
\Gamma ^{{\rm MSM}}(H\rightarrow W_{L}^{+}W_{L}^{-})
\vert _{{\rm tree}}
&=&\frac{1}{16\pi }\frac{m_{H}^{3}}{v^{2}}\sqrt{
1-\frac{4M_{W}^{2}}{m_{H}^{2}}}.
\end{eqnarray}
For $m_{H}=300$  GeV these formulae give 
$\Gamma ^{{\rm MSM}}(H\rightarrow Z_{L}Z_{L})=3.5
{\rm  GeV}$  and
\linebreak
\noindent  
$\Gamma ^{{\rm MSM}}(H\rightarrow W^{+}_{L}W^{-}
_{L})=7.5 {\rm  GeV}$. 
The tree level decay widths in the two Higgs doublet model 
are suppressed by a factor $\sin ^{2}(\alpha -\beta)$ 
compared with (51).
To take into account the one-loop corrections 
 in  MSM amounts to multiplying (51) by the factor
\begin{equation}
\left \vert 1+\frac{1}{4\pi ^{2}}\frac{m_{H}^{2}}
{v^{2}}\left (\frac{19}{16}-\frac{3\sqrt{3}\pi}{8}
+\frac{5\pi ^{2}}{48}\right )\right \vert ^{2}.
\end{equation}
This factor is about 1.013 for $m_{H}=300 {\rm GeV}$ 
and the radiative corrections have been said to be 
very small in MSM. 

The situation in the two-Higgs doublet model could differ 
very much from this.  The radiative corrections would
 be enhanced if $m_{G}^{2}/v^{2}$,  $m_{A}^{2}/v^{2}$ 
and   $m_{h}^{2}/v^{2}$ are large and could stand out 
from the lowest-order values. Our interest is to what 
extent they could supersede the tree-predictions.
Figs. 3 and Fig. 4 give us typical patterns of the 
$m_{G}$- and $m_{A}$-dependences. One can immediately 
notice that the decay rate of $H\rightarrow Z_{L}Z_{L}$ 
depends on $m_{G}$ and $m_{A}$ surprisingly  little in 
contrast to our naive expectation.
We have also confirmed that the $m_{h}$ dependence 
is  very weak. We can  hardly see any 
indication of power-behaviors w.r.t. heavier Higgs 
boson masses. 

The decay width of $H\rightarrow 
W^{+}_{L}W^{-}_{L}$ on the other hand exhibits some 
power-behavior. The one-loop corrections are potentially 
large to overcome  the tree-level 
predictions. One thing that we 
should pay attention here is that, while the radiative 
corrections are large in general, they  
tend to become small for $m_{G}=m_{A}$.
The apparent difference  lying between the two decay 
modes into $W$- and $Z$-pairs is  puzzling. If the isospin 
symmetry $SU(2)_{V}$ is exact, these 
two decay rates differ simply by 
a factor of two (up to the phase space difference) 
irrespectively of the choice of the parameters. In our 
two-Higgs doublet model, the isospin symmetry is broken 
by the $\lambda _{5}$-term in which we have necessarily 
to seek for the source of the difference. 

The overall properties of the decay widths may be 
seen easier if one would draw the widths in the birds' 
eye views. Figs 5 and 6 give   those of the
decay width of $\Gamma (H\rightarrow Z_{L}Z_{L})$ as 
a function of $m_{G}$ and $m_{A}$, and Figs. 7 and 8 
those of   $\Gamma (H\rightarrow W_{L}^{+}W_{L}^{-})$ .
Figs. 5 and 7 correspond to the case (I) and Figs. 6 and 8
 to the case (II).  

Figs. 5 and 6 show that the decay width of $H\rightarrow 
Z_{L}Z_{L}$ does not go far away from the tree-level 
values (3.5 GeV for Fig. 5 and 1.75 GeV for Fig. 6) for
a wide range of parameters,  $400 {\rm GeV} < m_{G}, m_{A}
 < 1000 {\rm GeV}$. This is rather in contrast with our 
naive expectation. Considering that $m_{G}^{2}/v^{2}$ 
and $m_{A}^{2}/v^{2}$ are greater than unity, the radiative 
corrections could instabilize the tree predictions. 
The reason for the almost flat behavior in Figs. 5 and 6
 will be elucidated qualitatively in Sec. 7.

In the case of $H\rightarrow W_{L}^{+}W_{L}^{-}$, 
on the other hand, we can immediately see in Fig. 7 
a conspicuous ridge elongated along the line
$m_{G}=m_{A}$. 
The width comes close to the tree level prediction
along this line, {\it i.e.},  
the radiative corrections tend to be 
minimized in the custodial $SU(2)_{V}$ symmetric limit.
The corrections become larger and larger as we go 
off  from this $m_{G}=m_{A}$ line. 
This reflects the fact that the radiative corrections 
entail the terms proportional to $m_{G}^{4}$, 
$m_{G}^{2}m_{A}^{2}$ and 
$m_{A}^{4}$ that exceed the tree-values.
 The shape  of the surface  in Fig. 8 corrsponding to the 
case II differs from that in Fig. 7. The one-loop   
predictions go up and down depending on the choice 
of mass parameters. The point  to be emphasized is that 
the line $m_{G}=m_{A}$ is again given a  special 
meaning. The radiative corrections  along 
this line tend to be  minimized, the tree-level width 
being equal to 3.8 GeV.

To sum up our  numerical computation, two questions have 
emerged naturally. One is why the radiative corrections to 
the decay rate of $H\rightarrow W_{L}^{+}W_{L}^{-}$  
are minimized for $m_{G}=m_{A}$. The other question 
is what explains the difference between 
the two decay rates and how.
We have included the top quark mass effect in Figs. 3-8. 
The top contributions, however, modify the predictions 
of the widths only within a few per cent and are not 
very significant.

\vskip1cm
\noindent 
{\large {\bf 7. New screening theorem for Higgs vertex}}
\vskip0.5cm

To shed light on the two questions raised at the 
end of  Sec. 6, we disentangle  the leading terms among 
${\cal M}_{Hww}(p^{2}=m_{H}^{2})$ and 
${\cal M}_{Hzz}(p^{2}=m_{H}^{2})$.  By leading terms we 
mean those containing  $m_{G}^{2}$, $m_{A}^{2}$, and 
$m_{h}^{2}$, but not $m_{H}^{2}$ at all. 
The results are as follows:
\begin{eqnarray}
{\cal M}_{Hww}(m_{H}^{2})\longrightarrow &  & \frac{1}
{(4\pi )^{2}v^{3}}\sin (\alpha -\beta )
  \left \{(m_{G}^{2}-m_{A}^{2})m_{A}^{2}
-m_{G}^{2}m_{A}^{2}\ln \frac{m_{G}^{2}}{m_{A}^{2}}\right \}
                          \nonumber \\
& &+({\rm  terms \; depending \; on \; the \;
 prescription \; for \; }
\delta \beta )
             \nonumber \\
& &+{\cal O}(\frac{m_{H}^{2}m_{G}^{2}}{v^{3}}, 
\frac{m_{H}^{2}m_{A}^{2}}{v^{3}}, \; {\rm or}\;
\frac{m_{H}^{2}m_{h}^{2}}{v^{3}}),  
\end{eqnarray}
\begin{equation}
{\cal M}_{Hzz}(m_{H}^{2})\longrightarrow 
\hskip0.5cm {\cal O}(
\frac{m_{H}^{2}m_{G}^{2}}{v^{3}}, 
\frac{m_{H}^{2}m_{A}^{2}}{v^{3}}, \; {\rm or}\;
\frac{m_{H}^{2}m_{h}^{2}}{v^{3}}).  
\end{equation}
The second line in (53) is due to the terms that arise by 
our choice of $\Pi _{zA}$ instead of $\Pi _{wG}$ 
in defining the renormalization $\delta \beta$, 
{\it i.e.}, Eq. (36).
In other words, these come from the last 
line of Eq. (50).

This leading behavior derived by hand in (53) and (54) 
is of course consistent 
with our numerical computations. 
The  decay into  a $W$-pair   
behaves as fourth power of $m_{G}$ and/or $m_{A}$, but 
does not contain $m_{h}$ in (53). It is therefore  
insensitive to $m_{h}$.  Furthermore the leading terms in 
(53) also disappear if we put $m_{G}=m_{A}$. Recall that 
the second  line in (53) consists of those 
terms containing $m_{G}^{2}-m_{A}^{2}$.
The behavior (54) indicates that all of the leading 
terms just disappear after summing up all the diagrams 
together with counterterms and the decay
rate $\Gamma (H\rightarrow Z_{L}Z_{L})$ is fairly
insensitive to $m_{G}$, $m_{A}$, or $m_{h}$.

From the viewpoint of the custodial $SU(2)_{V}$ 
  symmetry and its breaking,
 it is natural to divide the decay amplitudes into 
two parts
\begin{eqnarray}
{\cal M}_{Hww}&=&{\cal M}^{S}+{\cal M}^{B}_{Hww},
                       \\
{\cal M}_{Hzz}&=&{\cal M}^{S}+{\cal M}^{B}_{Hzz}.
\end{eqnarray}
Here ${\cal M}^{S}$ is the custodial symmetric part
and ${\cal M}^{B}_{Hww}$  and 
${\cal M}^{B}_{Hzz}$ are those due to the custodial 
symmetry  breaking. As we mentioned 
in Sec. 2, the custodial symmetry  breaking occurs 
through  the $\lambda _{5}$-coupling. This breaking 
effect shows up in two ways in the perturbative 
calculations. One is in the 
broken-symmetry in the Higgs-Goldstone  couplings, 
the other in the different masses in $G$ and $A$ 
propagators. Both of these two effects are collected 
in the second terms of (55) and (56).
The symmetric part ${\cal M}^{S}$ are obtained 
simply by replacing 
$m_{A}$ by $m_{G}$ everywhere in 
${\cal M}_{Hww}$ and ${\cal M}_{Hzz}$. 

The leading behavior shown in (53) and (54) indicates 
that the custodial symmetric part ${\cal M}^{S}$ does 
not possess the fourth-power  terms w.r.t. 
$m_{G}$, $m_{A}$ or $m_{h}$. We may conclude 
that ${\cal M}^{S}$ is insensitive to the heavier 
Higgs boson masses. We are thus led to postulate 
a new screening theorem.

\noindent
{\bf Theorem}:
\hskip0.3cm 
There occurs a cancellation mechanism among the 
leading terms w.r.t. the heavier Higgs boson masses 
in the custodial $SU(2)_{V}$-symmetric limit 
in the radiative corrections to the decay rates of 
$H\rightarrow Z_{L}Z_{L}$ and
$H\rightarrow W^{+}_{L}W^{-}_{L}$. 

It should be stressed  that this theorem 
differs from Veltman's in that the Veltman's theorem
has been proved for the oblique-type radiative 
corrections, while the above one is for the Higgs
 decay vertex. It should be also noticed that despite 
this difference the custodial  symmetry is the key
 ingredient in both cases. Recall that the proof 
by Einhorn and Wudka relies heavily on 
 the custodial  symmetry. 
What we did in Sec. 6 and in (53) and (54) amounts to
 confirming  the above theorem explicitly on the one-loop 
level by numerical calculations and by hand, 
respectively. We do not present here a general 
proof of our theorem  and we should not go 
so far as to say anything definite 
as to the two-loop and higher-order cases, 
only mentionimg the following:  
the fact that the custodial  symmetry is playing the 
key role in the Veltman's theorem   
is suggestive of the validity of our theorem beyond 
one-loop. It is also extremely tempting to speculate that 
there could occur a similar cancellation mechanism in 
 other decay modes of the $H$-boson. We will come to 
this important issue in our future publications.

Now let us move  on to the custodial symmetry 
 breaking part. The absence of the 
leading terms in ${\cal M}^{S}$, 
means that the sensitive properties 
 of the amplitude all come 
from ${\cal M}^{B}_{Hww}$ and ${\cal M}^{B}_{Hzz}$. 
In the $W$-pair dacay, ${\cal M}^{B}_{Hww}$ vanishes 
for $m_{G}=m_{A}$ that corresponds to the flat direction 
shown in Figs. 7 and 8.  Thus the answer to the first 
question raised in Sec. 6, namely, the minimization  
of the radiative corrections along the line $m_{G}=m_{A}$, 
 becomes almost self-evident 
once we accept our screening theorem.

The second question is what explains the apparent 
difference lying between decays into $W$- and $Z$-pairs.
As we remarked before, the 
custodial symmetry breaking effects
appear in the Higgs couplings on one hand, and the 
different $G$ and $A$ masses in the propagators on the 
other. Let us concentrate on the former, 
and in particular on the 
 broken-symmetry in the Higgs coupling 
triggered by the  $\lambda _{5}$-term  in the potential
\begin{eqnarray}
\lambda _{5}({\rm Im}\Phi _{1}^{\dag}\Phi _{2})^{2}
&=&-\frac{\lambda _{5}}{4}\Bigl [
(w^{+}G^{-}-G^{+}w^{-})   \Bigr.
                    \nonumber \\
& &+iA\left \{ v+h\cos (\alpha -\beta)-H\sin (
\alpha -\beta)\right \}
                  \nonumber \\
& &-iz\Bigl.  \left \{ 
h\sin (\alpha -\beta)+H\cos (\alpha -\beta )
\right \} \Bigr ]^{2}.
\end{eqnarray}
A close look at this expression shows  
that there exists an interaction 
$(w^{+}G^{-}-w^{-}G^{+})A$, 
while another possible term
\begin{equation}
(w^{+}G^{-}-w^{-}G^{+})z 
\end{equation}
is missing. The absence of this triple interaction
 indicates that there are some Feynman 
diagrams contributing to $H\rightarrow w^{+}w^{-}$ which 
do not have a conterpart in $H\rightarrow zz$. Actually
the $z$-interactions with charged scalars are peculiar:
not only (58) but also  terms such as 
\begin{equation}
w^{+}w^{-}z,\hskip0.5cm G^{+}G^{-}z, \hskip0.5cm
(w^{+}G^{-}+w^{-}G^{+})z,   
\end{equation}
are absent in the whole of the potential 
$V(\Phi _{1}, \Phi _{2})$. 
These are all forbidden interaction vertices if  
either G-parity or  {\it CP}-invariance is  exact.

This peculiarity leads to a considerable 
simplification in tracing the $m_{G}$ dependence 
in ${\cal M}_{Hzz}$ on 
which we now focus our attentions. 
Neither of $\Gamma _{Hzz}^{(1)}$ nor 
$\Gamma _{Hzz}^{(3)} $has the $m_{G}$-dependence at all, 
as one can make sure of  in Eqs. (41) and (43).
This is simply because there is no Feynman diagram of 
the type of Fig. 2(a) or Fig. 2(c) involving $G^{\pm}$.
In the heavy Higgs mass limit 
($m_{G}, m_{A}, m_{h}\gg m_{H}$ ),
 the $m_{G}$-dependence of the 
amplitude of the $Z_{L}Z_{L}$-decay is governed by the 
combination
\begin{equation}
\Gamma _{Hzz}^{(2)}+\frac{1}{v}\cos (\alpha -\beta)
{\rm Re}(\Pi _{hH}(m_{H}^{2}))-\frac{1}{v}\sin (\alpha 
-\beta){\rm Re}(\Pi _{HH}(m_{H}^{2})).
\end{equation}
All the other terms in (49) are sub-leading 
in the heavy Higgs mass limit . 
The $m_{G}$-dependence comes about from the Feynman 
diagrams, 
Fig. 1(a) and Fig. 2(b) in which either of 
$wG$,  or $GG$ pair is encircling. 

Now notice that the last two terms in (60) 
join together to produce  the two-point function  
connecting the state $\vert H >$ with the linear 
combination
\begin{equation}
\frac{1}{v}\cos (\alpha -\beta)\vert h >-\frac{1}{v}
\sin (\alpha -\beta)\vert H >.
\end{equation} 
An important point to be observed here is that the
 couplings of the combination (61) to $wG$- 
and $GG$-pairs 
are exactly of the same strength with opposite sign 
as the quartic $zz(w^{+}G^{-}+w^{-}G^{+})$ and 
$zzG^{+}G^{-}$ couplings, respectively, that contribute 
to $\Gamma _{Hzz}^{(2)}$.
Thus the leading terms containing $m_{G}$
({\it i.e.}, $m_{G}^{4}/v^{3}$, $m_{G}^{2}m_{A}^{2}/v^{3}$
 and $m_{G}^{2}m_{h}^{2}/v^{3}$) all 
disappear in (60) and thus in ${\cal M}_{Hzz}$ as well. 

The equality  of the strength of quartic
 and triple couplings as 
described  above may be undestood easily if we 
go over to the so-called Georgi-basis [32], in which  
the combinations
\begin{eqnarray}
& &v+h\cos (\alpha -\beta)-H\sin (\alpha -\beta)+iz,
                           \nonumber \\
& &h\sin (\alpha -\beta)+H\cos  (\alpha -\beta)+iA
\end{eqnarray}
are taken  from the outset. It is also to be 
mentioned that the equality  of the quartic and triple 
couplings is necessary for the ultraviolet divergences 
to disappear in the combination (60).

Once the absence of leading terms of the form 
$m_{G}^{4}/v^{3}$,    
$m_{G}^{2}m_{h}^{2}/v^{3}$ and   
$m_{G}^{2}m_{A}^{2}/v^{3}$ 
is   established in ${\cal M}_{Hzz}^{B}$, then  
it is almost trivial that those proportional to 
$m_{A}^{4}/v^{3}$   and 
$m_{A}^{2}m_{h}^{2}/v^{3}$ are also absent. This is 
because    the leading terms in 
${\cal M}_{Hzz}^{B}$ must vanish by setting 
$m_{G}=m_{A}$ and this is possible only when those terms 
are  not there. 

In this way we are able to  disentangle so many Feynman 
diagrams and to trace the origin of the difference 
lying between ${\cal M}_{Hww}$ and ${\cal M}_{Hzz}$.
We are now convinced how and why the moderate behaviors of 
$\Gamma (H\rightarrow Z_{L}Z_{L})$ are seen in 
Figs. 5 and 6.

\vskip1cm
\noindent 
{\large {\bf 8.  Summary and Discussions}}
\vskip0.5cm

In the present paper, we have studied one-loop radiative 
corrections to the Higgs boson ($H$)
 decay into  longitudinal 
gauge boson pairs,    $H\rightarrow Z_{L}Z_{L}$ 
 and $H\rightarrow W^{+}_{L}W^{-}_{L}$. The two Higgs 
doublet model has been considered throughout and $H$ 
is assumed to be much heavier than the weak gauge bosons. 
A particular attention has been paid to the 
non-decoupling effects due to the other  Higgs bosons 
which are assumed to be all heavier than $H$. 
 
Our numerical analyses on the decay rate 
$H\rightarrow W_{L}^{+}W_{L}^{-}$ show  that the 
radiative corrections are potentially large.
The larger the mass difference $m_{G}^{2}-m_{A}^{2}$ 
becomes, the larger the deviation 
from the tree-predictions 
turns out to be. The point is that the radiative corrections 
tend to be minimized if $m_{G}^{2}=m_{A}^{2}$, for which 
the custodial $SU(2)_{V}$-symmetry is recovered.
This fact has led us to postulate a new screening theorem 
for Higgs vertices with reference to the custodial 
symmetry, which may be regarded as a generalization of the 
Veltman's theorem. 
 
Although our generalized screening theorem has been 
confirmed only at the one-loop level for the particular 
decay processes, this type of theorem 
 would hopefully play the role of a  
working-hypothesis in our future study. It will be of 
particular interest to see if the custodial symmetry
would have any relevance to the screening effect
 of heavy particles in other types of non-decoupling
 processes. Some examples to be examined 
include triple gauge boson couplings [33] and longitudinal 
gauge boson scatterings, 
$W^{+}_{L}W^{-}_{L}\rightarrow W^{+}_{L}W^{-}_{L}$  and
$W^{+}_{L}W^{-}_{L}\rightarrow Z_{L}Z_{L}$. 
It is also important to see what happens in models other 
than the two-Higgs doublet model.
We will come to these subjects  in our future publications.

The decay rate $\Gamma (H\rightarrow Z_{L}Z_{L})$ has turned 
out to be unexpectedly insensitive to heavier Higgs 
boson masses. The reason for this  was elucidated in detail 
in Sec. 7. It is thus rather difficult to use this decay 
rate to get any signature 
of  the masses of unknown heavy particles. Alternatively, 
however, by turning the tables around the decay 
$H \rightarrow Z_{L}Z_{L}$ could be useful  to get 
information on the mixing angle $\alpha$. Fig. 9 shows the 
$\alpha $-dependence of this decay rate for 
$\tan \beta =2$, $\tan \beta =5$ and $\tan \beta =10$. 
Measurements of the width with  an accuracy on the order 
of a fraction of 1 GeV would enable us to determine  the 
value of $\alpha $.

Finally we would like to add a few rather peripheral 
remarks. As we mentioned in Sec. 2, we neglected the term 
$(\mu _{12}^{2}\Phi _{1}^{\dag }\Phi _{2}+
\mu _{12}^{2*}\Phi _{2}^{\dag }\Phi _{1})$ 
 which would break the 
discrete symmetry of the Higgs potential (1).  
By dropping this term, the  heavy 
Higgs boson mass limit is rendered to be  the same as
 the strong quartic couplings (see Eqs. (7)-(11)).
Thus the non-decoupling effects are expected to be 
potentially large because of the strong couplings.
If we would  included 
$(\mu _{12}^{2}\Phi _{1}^{\dag }\Phi _{2}+
\mu _{12}^{2*}\Phi _{2}^{\dag }\Phi _{1})$ as in 
minimally supersymmetric models, one may wonder whether 
 the non-decoupling effects could be expected or not. 
In the presence of the mass scale $\mu _{12}$ in addition 
to $v$, the heavy Higgs mass limit does not always imply 
the strong quartic couplings. There exists a limit in which 
$\mu _{12}$, $m_{h}$, $m_{A}$ and $m_{G}$ are all large 
while the quartic coulings are small. In such a case, 
the decoupling is  expected 
from the begining and the two Higgs doublet model becomes  
similar to the minimal standard model 
at low energies [34]. 

The non-decoupling effects considered in this paper may be
studied by the electro-weak chiral Lagrangian approach 
[35-37]. This method is powerful in the minimal standard 
model for systematic studies of low-energy manifestation 
of heavy particles. Whether this method is useful in
 the two Higgs doublet model  as well is yet to be 
scrutinized  and we leave it as an open  problem.

\vskip2cm
\noindent
{\large
{\bf Acknowledgements}
}

\vskip0.5cm
\noindent
We  would like to thank 
M.M.   Nojiri, M. Tanabashi and M. Tanaka for 
stimulating discussions, and M. Peskin for useful 
suggestions.  This work is supported in part by the Grant 
in Aid for Scientific Research from the Ministry of 
Education, Science and Culture (no.  06640396). 
\vskip2cm
\vfill\eject 
\noindent
{\bf Appendix A}

\vskip0.5cm
The Yukawa couplings of top and bottom quarks  are
 given by
\begin{equation}
{\cal L}_{{\rm Yukawa}}=-\frac{\sqrt{2}m_{b}}
{v\sin \beta}{\bar Q}_{L}\Phi _{2}b_{R}
-\frac{\sqrt{2}m_{t}}{v\sin \beta}{\bar Q}_{L}
i\tau _{2}\Phi _{2}^{*}t_{R}+{\rm h.c.},
\end{equation}
for model I and 
\begin{equation}
{\cal L}_{{\rm Yukawa}}=-\frac{\sqrt{2}m_{b}}
{v\cos  \beta}{\bar Q}_{L}\Phi _{1}b_{R}
-\frac{\sqrt{2}m_{t}}{v\sin \beta}{\bar Q}_{L}i\tau _{2}
\Phi _{2}^{*}t_{R}+{\rm h.c.},
\end{equation}
for model II, 
where $Q_{L}=(t_{L}, b_{L})^{T}$. More explicitly, 
these interactions are expanded by putting Eqs. (3)-(6) 
as follows;
\begin{eqnarray}
{\cal L}_{{\rm Yukawa}}
&=&-m_{b}{\bar b}b+C_{1}{\bar b}bH
+C_{2}{\bar b}bh+C_{3}{\bar b}i
\gamma _{5}bz+C_{4}{\bar b}i\gamma _{5}bA
           \nonumber \\
& &-m_{t}{\bar t}t+D_{1}{\bar t}tH
+D_{2}{\bar t}th+D_{3}{\bar t}i
\gamma _{5}tz+D_{4}{\bar t}i\gamma _{5}tA
           \nonumber \\
& &+E_{1}\left \{{\bar t}(1+\gamma _{5})bw^{+}
+{\bar b}(1-\gamma _{5})tw^{-} \right \}   
             \nonumber \\
& &+E_{2}\left \{{\bar t}(1+\gamma _{5})bG^{+}
+{\bar b}(1-\gamma _{5})tG^{-} \right \}
                                  \nonumber \\
& &+F_{1}\left  \{{\bar t}(1-\gamma _{5})bw^{+}
+{\bar b}(1+\gamma _{5})tw^{-} \right \}   
                      \nonumber \\
& &+F_{2}\left \{{\bar t}(1-\gamma _{5})bG^{+}
+{\bar b}(1+\gamma _{5})tG^{-}\right \}.   
\end{eqnarray}
The coefficients in the above are tabulated in 
Table 1.

\vskip1cm
\vfill\eject
\noindent
{\bf Appendix B}
\vskip0.5cm

We define the following functions to express 
the various Feynman integrals:
\begin{eqnarray}
f_{1}(m^{2})&=& \mu ^{4-D} 
\int \frac{d^{D}k}{(2\pi )^{D}}\frac{i}
{k^{2}-m^{2}}    \nonumber             \\
&=&\frac{m^{2}}{(4\pi )^{2}}\left (\frac{2}{D-4}-1+
\gamma _{E}+\ln \frac{m^{2}}{4\pi \mu ^{2}}\right ),
               \nonumber \\ 
\end{eqnarray}
\begin{equation}
f_{2}(p^{2},m_{1}^{2},m_{2}^{2})=-i\mu ^{4-D}\int \frac{
d^{D}k}{(2\pi )^{D}}\frac{i}{k^{2}-m_{1}^{2}}\frac{i}
{(p-k)^{2}-m_{2}^{2}},    
\end{equation}

\begin{equation}
g(p^{2},  m_{1}^{2},  m_{2}^{2},  m_{3}^{2})=
\int \frac{d^{4}k}{(2\pi )^{4}}\frac{i}{(k-p_{1})^{2}-
m_{1}^{2}}\frac{i}{(k+p_{2})^{2}-m_{2}^{2}}
\frac{i}{k^{2}-m_{3}^{2}}.
\end{equation}
The function (68) can be expressed in terms of some 
combinations of the Spence function. Details are 
given in Appendix A of Ref. [5].

\vskip1cm
\noindent
{\bf Appendix C}
\vskip0.5cm

Here  we would like to summarize the full expressions 
of the various two-point functions used in the text.
Let us begin with $\Pi _{zA}(p^{2})$ and 
$\Pi _{zz}(p^{2})$, which are both 
vanishing at $p^{2}=0$ as a reflection 
of the Nambu-Goldstone 
theorem and are put in the following forms
\begin{equation}
\Pi _{zz}(p^{2})={\hat \Pi} _{zz}(p^{2})-{\hat \Pi}
 _{zz}(0),
\hskip1cm
\Pi _{zA}(p^{2})={\hat \Pi} _{zA}(p^{2})-{\hat \Pi}
 _{zA}(0).
\end{equation}
The functions 
${\hat \Pi } _{zz}(p^{2})$ and 
${\hat \Pi } _{zA}(p^{2})$ are given as a sum of scalar 
and quark contributions 
\begin{equation}
{\hat \Pi} _{zz}(p^{2})={\hat \Pi} ^{{\rm (scalar)}}_{zz}
(p^{2})+{\hat \Pi} ^{({\rm quark})}_{zz}(p^{2}),
\hskip0.5cm
{\hat \Pi} _{zA}(p^{2})={\hat \Pi} ^{({\rm scalar})}_{zA}
(p^{2})+{\hat \Pi} ^{({\rm quark})}_{zA}(p^{2})
\end{equation}
and each of these terms is  given as follows:
\begin{eqnarray}
{\hat \Pi }^{({\rm scalar})}_{zz}(p^{2})&=&
\frac{m_{h}^{4}}{v^{2}}\cos ^{2}(
\alpha -\beta)f_{2}(p^{2},m_{h}^{2},0)
               \nonumber \\
& &+\frac{m_{H}^{4}}{v^{2}}\sin  ^{2}(
\alpha -\beta)f_{2}(p^{2},m_{H}^{2},0)
               \nonumber \\
& &+\frac{(m_{h}^{2}-m_{A}^{2})^{2}}{v^{2}}\sin ^{2}
(\alpha -\beta)f_{2}(p^{2},m_{h}^{2},m_{A}^{2})
               \nonumber \\
& &+\frac{(m_{H}^{2}-m_{A}^{2})^{2}}{v^{2}}\cos ^{2}
(\alpha -\beta)f_{2}(p^{2},m_{H}^{2},m_{A}^{2}),
               \nonumber \\
\end{eqnarray}
\begin{equation}
{\hat \Pi }^{({\rm quark})}_{zz}(p^{2})=
2N_{C}\frac{m_{t}^{2}}{v^{2}}
f_{2}(p^{2}, m_{t}^{2},  m_{t}^{2})p^{2},
\end{equation}
\begin{eqnarray}
{\hat \Pi }^{({\rm scalar})}_{zA}(p^{2})   
&=&\frac{m_{h}^{2}(m_{h}^{2}-m_{A}^{2})}{2v^{2}}
\sin (2\alpha -2\beta)f_{2}(p^{2}, m_{h}^{2}, 0) 
                  \nonumber \\
& &-\frac{m_{H}^{2}(m_{H}^{2}-m_{A}^{2})}{2v^{2}}
\sin (2\alpha -2\beta)f_{2}(p^{2}, m_{H}^{2}, 0)     
                       \nonumber \\
& &+\frac{m_{h}^{2}-m_{A}^{2}}{v^{2}}\sin (\alpha -\beta)
\{m_{h}^{2}(\frac{\sin ^{2}\beta \cos \alpha }
{\cos \beta}+\frac{\cos ^{2}\beta \sin \alpha}
{\sin \beta})   \nonumber \\
& &+2m_{A}^{2}\cos (\alpha -\beta)\}
f_{2}(p^{2}, m_{h}^{2},m_{A}^{2})
                  \nonumber \\
& &+\frac{m_{H}^{2}-m_{A}^{2}}{v^{2}}\cos (\alpha -\beta)
\{m_{H}^{2}(\frac{\cos ^{2}\beta \cos \alpha }
{\sin \beta}-\frac{\sin ^{2}\beta \sin \alpha}
{\cos \beta})      \nonumber \\
& &-2m_{A}^{2}\sin (\alpha -\beta)\}
f_{2}(p^{2}, m_{H}^{2},m_{A}^{2}),
                  \nonumber \\
\end{eqnarray}
\begin{equation}
{\hat \Pi }^{({\rm quark})}_{zA}(p^{2})=
2N_{C}\frac{m_{t}^{2}}{v^{2}}\cot \beta 
f_{2}(p^{2}, m_{t}^{2}, m_{t}^{2})p^{2}.
\end{equation}

Similarly, $\Pi _{ww}(p^{2})$ and 
$\Pi _{wG}(p^{2})$ are given in the same form as in Eqs. 
(69) and (70).  
${\hat \Pi }^{({\rm scalar})}_{ww}(p^{2})$ and 
${\hat \Pi }^{({\rm scalar})}_{wG}(p^{2})$
 were give explicitly 
in Ref. [5] and the top quark contributions are 
\begin{equation}
{\hat \Pi }^{({\rm quark})}_{ww}(p^{2})=-2N_{C}
\frac{m_{t}^{2}}{v^{2}}(m_{t}^{2}-p^{2})f_{2}
(p^{2}, m_{t}^{2}, 0),
\end{equation}
\begin{equation}
{\hat \Pi }^{({\rm quark})}_{wG}(p^{2})=-2N_{C}
\frac{m_{t}^{2}}{v^{2}}\cot \beta (m_{t}^{2}-p^{2})f_{2}
(p^{2}, m_{t}^{2}, 0).
\end{equation}

\vskip1cm
\vfill\eject
\noindent
{\bf Appendix D}
\vskip0.5cm

The other two-point functions 
$\Pi _{HH}(p^{2})$ and $\Pi _{hH}(p^{2})$ were given in 
Appendices C and D in  Ref. [5]
 except for the quark contributions, 
\begin{eqnarray}
\Pi _{HH}^{({\rm quark})}&=&
-2N_{C}\frac{m_{t}^{2}}{v^{2}}
\frac{\cos ^{2}\alpha }{\sin ^{2}\beta }
\left \{2f_{1}(m_{t}^{2})+(4m_{t}^{2}-p^{2})f_{2}(p^{2}, 
m_{t}^{2},  m_{t}^{2}) \right \}
                       \nonumber \\
& &+\frac{N_{C}}{v}\left (\frac{\cos ^{3}\alpha}
{\sin \beta}-\frac{\sin ^{3}\alpha}{\cos \beta}\right )
T_{H}^{({\rm quark})}
+\frac{N_{C}}{v}\frac{\sin 2\alpha }{\sin 2\beta }
\cos (\alpha -\beta )T_{h}^{({\rm quark})},
                       \nonumber \\
\end{eqnarray}
\begin{eqnarray}
\Pi _{hH}^{({\rm quark})}&=&
-2N_{C}\frac{m_{t}^{2}}{v^{2}}
\frac{\sin \alpha \cos \alpha}{\sin ^{2}\beta}
\left \{2f_{1}(m_{t}^{2})+(4m_{t}^{2}-p^{2})
f_{2}(p^{2}, m_{t}^{2},  m_{t}^{2})\right \}
                             \nonumber \\
& &+\frac{N_{C}}{v}\frac{\sin 2\alpha }
{\sin 2\beta }\left \{\cos (\alpha-\beta)
T_{H}^{({\rm quark})} +\sin (
\alpha -\beta)T_{h}^{({\rm quark})}\right \}. 
                            \nonumber \\
\end{eqnarray}
Here the tad pole effects on  the $H$ and $h$ fields 
due to the top quark are denoted by 
$T^{({\rm quark})}_{H}$ and 
$T^{({\rm quark})}_{h}$ respectively. These are given by
\begin{equation}
T_{H}^{({\rm quark})}=
4\frac{m_{t}^{2}}{v}\frac{\cos \alpha }{\sin \beta }
f_{1}(m_{t}^{2}),
\end{equation}
\begin{equation}
T_{h}^{({\rm quark})}=
4\frac{m_{t}^{2}}{v}\frac{\sin \alpha }{\sin \beta }
f_{1}(m_{t}^{2}).
\end{equation}

\pagebreak
\noindent
{\bf References}

\begin{description}
\item{[1]}
F. Abe et al. (CDF Collaboration), 
Phys. Rev. Lett. {\bf 73}  (1994) 225; 
Phys. Rev. {\bf D 50} (1994) 2966.
\item{[2]}
F. Abe et al. (CDF Collaboration), 
Phys. Rev. Lett. {\bf 74} (1995) 2626;
S. Abachi et al. (D0 Collaboration), 
Phys. Rev. Lett. {\bf 74} (1995) 2632.
\item{[3]}
J.F. Gunion, H.E. Haber, G. Kane and S. Dawson, 
{\it The Higgs Hunter's Guide }
(Addison Wesley Publishing
 Company, 1990))
\item{[4]} 
M. Veltman, Acta. Phys. Pol. {\bf B 8}    
 (1977) 475; {\it ibidem} {\bf B 25} (1994) 1627;
  Phys. Lett. {\bf 70 B}    (1977) 253.    
\item{[5]}
S. Kanemura and T. Kubota, Phys. Rev. {\bf D 51}  
 (1995) 2462.
\item{[6]}
M. Veltman,  Nucl. Phys. {\bf B 123} (1977) 89;
P.  Sikivie,  L.  Susskind, M.  Voloshin and V.  
Zakharov,  Nucl.  Phys.  {\bf B 173}  (1980) 189;
M.B. Einhorn,D.R.T. Jones and M. Veltman,  
Nucl. Phys.  {\bf B 191} (1981) 146. 
\item{[7]} M. Einhorn and J. Wudka, Phys. Rev. 
{\bf D 39} (1989) 2758; {\it ibidem} {\bf D 47}
  (1993) 5029. 
\item{[8]}
S. Dawson and S. Willenbrock, Phys. Lett. {\bf B 211} 
(1988) 200; M.S. Chanowitz, M.A. Furman and I. Hinchliffe, 
Nucl. Phys. {\bf B 153} (1979) 402. 
\item{[9]}
S. Dawson and S. Willenbrock,  Phys. Rev. Lett. 
{\bf 62} (1989) 1232; Phys. Rev. {\bf D 40} 
(1989) 2880;
W.J. Marciano and S.S.D. Willenbrock, Phys. Rev. 
{\bf D 37} (1988) 2509.
\item{[10]}
B.A. Kniehl,  Nucl. Phys. {\bf B 357} (1991) 439; 
Nucl. Phys. {\bf B 352} (1991) 1.
\item{[11]}
D. Pierce and A. Papadopoulos,   Phys. Rev. {\bf D 47}
 (1993) 222.
\item{[12]}
S.N. Gupta, J.M. Johnson  and  W.W. Repko, 
Phys. Rev. {\bf D 48} (1993) 2083;
L. Durand,  J.M. Johnson and J.L. Lopez, Phys. Rev. 
{\bf D 45} (1992) 3112; 
P.N. Maher, L. Durand  and K. Riesselmann,
  Phys. Rev. {\bf D 48} (1993) 1061. 
\item{[13]}
S. Kanemura, T. Kubota and H. Tohyama, OU-HET 233
(hep-ph/9601281), to appear in Proceedings of {\it 
``Workshop on Physics and Experiments with Linear 
Colliders"} (  World Scientific Pub. 
Co. )
\item{[14]}
J.M. Cornwall, D.N. Levin and  G. Tiktopoulos,
  Phys. Rev.  Lett. {\bf 30} (1973) 1268; Phys. 
  Rev.  {\bf D 10} (1974) 1145. 
\item{[15]}
B.W. Lee, C. Quigg and H.B. Thacker, Phys. Rev.
{\bf D 16} (1977) 1519.
\item{[16]}
M.S. Chanowitz and M.K. Gaillard, Nucl. Phys. 
{\bf B 261} (1985) 379;
G. Gounaris,  R. K{\" o}gerler and H. Neufeld, 
Phys. Rev. {\bf D 34} (1986) 3257;
 Y.P. Yao and C.P. Yuan, Phys. 
Rev. {\bf D 38} (1988) 2237; H. Veltman, Phys. Rev. 
{\bf D 41} (1990) 2294; J. Bagger and C Schmit, Phys. 
Rev. {\bf D 41} (1990) 264;  K Aoki, {\it Proceedings 
of the Meeting on Physics at TeV Energy Scale} (KEK 
Report 89-20, Nov. 1987) p.  20.
\item{[17]}
H.-J. He,  Y.-P. Kuang and X. Li, Phys. Rev. Lett. 
{\bf 69} (1992) 2619; Phys. Rev. {\bf D 49} 
(1994) 4842.
\item{[18]}
S.L. Glashow and S. Weinberg, Phys. Rev. {\bf D 15} 
 (1977) 1958;
J. Liu and L. Wolfenstein,  Nucl. Phys. {\bf B 289}
 (1987) 1.
\item{[19]}
L. McLerran, M. Shaposhnikov,  N. Turok and M. Voloshin, 
Nucl. Phys. {\bf B 256} (1991) 451;  
A.G. Cohen, D.B. Kaplan and A.E. Nelson,  Phys. Lett.
 {\bf B 263} (1991) 86; 
A.E. Nelson, D.B. Kaplan and A.G. Cohen,  Nucl. Phys.
 {\bf B373} (1992) 453;
A.G. Cohen, D.B. Kaplan and A.E. Nelson, 
Ann. Rev. of Nucl. and Part.  Sci. {\bf 43} (1993) 27.
\item{[20]} 
D. Toussaint, Phys. Rev. {\bf D 18} (1977) 1626.
\item {[21]}
H.E. Haber, G.L. Kane and T. Stirling, 
Nucl. Phys. {\bf B 161} (1979) 493.
\item{[22]}
J.F. Donoghue and L.F. Li, Phys. Rev. {\bf D 19} 
(1979) 945; L. Hall and M. Wise, Nucl. Phys. {\bf B187} 
(1981) 397.
\item{[23]}
K. Aoki, Z. Hioki, R. Kawabe, M. Konuma and T. Muta, 
Prog. Theor. Phys. Suppl. {\bf 73} (1982) 1;
M. B{\" o}hm, W. Hollik and H. Spiesberger, Fortscr. Phys. 
{\bf 34} (1986) 687; 
W.F.L. Hollik, Fortschr. Phys. {\bf 38} (1990) 165.
\item{[24]}
A. Sirlin, Phys. Rev. {\bf D 22} (1980) 971;
W.J. Marciano and A. Sirlin, Phys. Rev. {\bf D 22}
(1980) 2695, [Erratum: {\bf D 31} (1985) 213];
A. Sirlin and W.J. Marciano, Nucl. Phys. {\bf B 189}
 (1981) 442.
\item{[25]}
S. Bertolini,  Nucl.  Phys. {\bf B 272} (1986) 77; 
W. Hollik,  Z. Phys.  {\bf C 32} (1986) 291; 
{\it ibidem} {\bf C 37} (1988) 569;
A. Denner, R.J. Guth and J.H. K{\" u}hn,  Phys. Lett. 
{\bf B 240} (1990) 438;
T. Inami, C.S. Lim and A. Yamada,  Mod. Phys. Lett.
  {\bf A7} (1992) 2789;
C.D. Froggatt, R.G. Moorhouse and I.G. Knowles,  
Phys. Rev. {\bf D 45} (1992) 2471.
\item{[26]}
V. Barger,  J.L.  Hewett and R.J.N. Phillips,  
Phys. Rev.  {\bf D 41} (1990) 3421;
A. J. Buras, P.  Krawczyk,  M.E. Lautenbacher 
and C.  Salazar,  Nucl. Phys.  {\bf B 337} (1990) 284;
J.F. Gunion and B. Grzadkowski,  Phys. 
Lett. {\bf B 243} (1990) 301;  
D. Cocolicchio and J.-R. Cudell,  Phys. Lett.
 {\bf B245} (1990) 591;
J. Rosiek,  Phys. Lett. {\bf B 252} (1990) 135;
G. Cvetic, preprint DO-TH 94/27 (November 1994).
\item{[27]}
B. Barish et al.(CLEO Collaboration), preprint 
CLEO CONF 94-1;F. Bartelt et al.(CLEO Collaboration),
Phys. Rev. Lett. {\bf 71} (1993) 4111; M.S. Alam et al.
(CLEO Collaboration), Phys. Rev. Lett. 
{\bf 71} (1993) 674;{\it ibidem} {\bf 74} (1995) 2885.  
\item{[28]}
K. Hagiwara, Talk presented at {\it XVII 
International Symposium on Lepton and Photon 
Interactions at High Energies} (10-15 August, 1995, 
Beijing, China) (hep-ph/9512425) ;
D. Atwood, L. Reina and A. Soni, preprint 
CEBAF-TH-96-01 (hep-ph/9603210).
\item{[29]}
A.K. Grant,  Phys. Rev. {\bf D 51} (1995) 207.   
\item{[30]}
D. Kominis and R.S. Chivukula,  Phys. Lett. 
{\bf B 304} (1993) 152.
\item{[31]}
S. Kanemura, T. Kubota and E. Takasugi,  Phys. 
Lett. {\bf B 313} (1993) 155; 
J.  Maalampi, J.  Sirkka and I.  Vilja,
  Phys. Lett. {\bf B 265} (1991) 371. 
\item{[32]}
H. Georgi,  Hadr. J.  Phys.  {\bf 1} (1978) 155.
\item{[33]}
K. Hagiwara, K. Hikasa, R. Peccei and D. Zeppenfeld, 
Nucl. Phys. {\bf B 282} (1987) 253;
C. Ahn, M.E. Peskin, B.W. Lynn and S. Selipsky, Nucl. 
Phys. {\bf B 309} (1988) 221; T. Appelquist and 
G.-H. Wu, Phys. Rev. {\bf D 48} (1993) 3235;
{\it ibidem } {\bf D 51} (1995) 240;
T. Inami, C.S. Lim, B. Takeuchi and M. Tanabashi, 
preprint (July, 1995, hep-ph/9510368)
\item{[34]}
C.S. Lim, T. Inami and N. Sakai, Phys. Rev. {\bf D 29} 
(1984) 1488;
L. Alvarez-Gaume, J. Polchinski and M.B. Wise,
 Nucl. Phys.  {\bf B 221} (1983) 495;
J.F. Gunion and A. Turski, Phys. Rev. {\bf D 39} 
(1989) 2701; 
{\it ibidem} {\bf D 40} (1989) 2325, 2333.
\item{[35]}
S. Weinberg, Physica {\bf 96 A} (1979) 327;
 J. Gasser and H. Leutwyler, Ann. Phys. (N.Y.) {\bf 158} 
(1984) 142; Nucl. Phys. {\bf B 250} (1985) 465.
\item{[36]}
T. Appelquist and C. Bernard, Phys. Rev. {\bf D 22} 
(1980) 200;     A.C. Longuitano, Nucl. Phys. {\bf B 188} 
(1981) 118; Phys. Rev. {\bf D 22} (1980) 1166.
\item{[37]}
M.J. Herrero and E.R. Morales, Nucl Phys. {\bf B 418} 
(1994) 431; {\it ibidem} {\bf B 437} (1995) 319.
\end{description}
\vfill\eject
\begin{flushleft}
{\bf Table 1.}
\end{flushleft}
\noindent
Yukawa couplings in  (65).
\begin{center}
\begin{tabular}{c|c|c}           \hline\hline 
Coefficients  &   Model I   & Model II     \\   \hline
$C_{1}$   & $-(m_{b}/v)\cos \alpha /\sin \beta $ 
          & $(m_{b}/v)\sin \alpha /\cos \beta $  \\
$C_{2}$   & $-(m_{b}/v)\sin \alpha /\sin \beta $ 
          & $-(m_{b}/v)\cos \alpha /\cos \beta $ \\
$C_{3}$   & $-m_{b}/v$ 
          & $-m_{b}/v$   \\
$C_{4}$   & $-(m_{b}/v)\cot \beta$ 
          & $ (m_{b}/v)\tan \beta $   \\
\hline
$D_{1}$   & $-(m_{t}/v)\cos \alpha /\sin \beta$ 
          & $-(m_{t}/v)\cos \alpha /\sin \beta$  \\
$D_{2}$   & $-(m_{t}/v)\sin \alpha /\sin \beta$ 
          & $-(m_{t}/v)\sin \alpha /\sin \beta$ \\
$D_{3}$   & $m_{t}/v$ 
          & $m_{t}/v$   \\
$D_{4}$   & $(m_{t}/v)\cot \beta $
          & $(m_{t}/v)\cot \beta   $   \\
\hline
$E_{1}$   & $-m_{b}/\sqrt{2}v$ 
          & $-m_{b}/\sqrt{2}v$      \\
$E_{2}$   & $-(m_{b}/\sqrt{2}v)\cot \beta$  
          & $(m_{b}/\sqrt{2}v)\tan \beta $   \\
$F_{1}$   & $m_{t}/\sqrt{2}v$ 
          & $m_{t}/\sqrt{2}v$      \\
$F
_{2}$   & $(m_{t}/\sqrt{2}v)\cot \beta$  
          & $(m_{t}/\sqrt{2}v)\cot \beta $   \\
\hline\hline
\end{tabular}
\end{center}
\vfill\eject
\begin{flushleft}
{\bf Table 2.}
\end{flushleft}
\noindent
Combinations of internal particles $(X, Y)$ 
running  
in Fig. 1(a) and $X$ in Fig. 1(b) 
contributing to $\Pi _{zz}$ and $\Pi _{zA}$ .
\begin{center}
\begin{tabular}{c|c|l}       \hline\hline
Propagator  & type & Internal particle species
                                  \\  \hline
$ \Pi _{zz}$     & Fig. 1(a) &    
     $(H, z)$,  $(H, A)$,  $(h, z)$,  $(h, A)$, 
     $( t,  {\bar t})$  
                                              \\
                 & Fig. 1(b) &
     $G$,  $A$, $h$, $H$
                                     \\ \hline
$\Pi _{zA}$     & Fig. 1(a) &    
     $(H, z)$,  $(H, A)$,  $(h, z)$,  $(h, A)$,
     $(t,  {\bar t})$  
                                              \\
                & Fig. 1(b) &   
     $G$,  $A$, $h$, $H$                      \\
                               \hline\hline
\end{tabular}
\end{center}
\vskip3cm

\begin{flushleft}
{\bf Table 3.}
\end{flushleft}
Combinations of internal particles (X, Y; Z) in Fig. 2(a) 
and (X, Y) in Figs. 2(b) and 2(c) for the vertices 
$\Gamma _{Hzz}^{(i)}$ ($i=1,2,3$) and $\Gamma _{Hzz}
^{{\rm (quark)}}$.
\begin{center}
\begin{tabular}{c|c|l}           \hline\hline 
Vertex  &   type & Internal particle species 
  \\   \hline
$\Gamma _{Hzz}^{(1)}$   & Fig. 2(a) &
          $(z,z;h)$,  $(z,z;H)$,  $(A,A;h)$,  $(A,A;H)$,  \\
  &  &    $(A,z;H)$,  $(A,z;h)$,  $(H,H;z)$,  $(H,H;A)$,  \\
  &  &    $(H,h;z)$,  $(H,h;A)$,  $(h,h;z)$,  $(h,h;A)$,  \\
$\Gamma _{Hzz}^{(2)}$   &  Fig. 2(b) &
       $(w,w)$,  $(z,z)$,  $(G,G)$,  $(A,A)$,  $(A,z)$,\\
  & &    $(G,w)$,  $(H,H)$,  $(H,h)$,  $(h,h)$          \\
$\Gamma _{Hzz}^{(3)}$    & Fig. 2(c) &
       $(z,h)$,  $(z,H)$,  $(A,h)$,  $(A,H)$,  \\
$\Gamma _{Hzz}^{\rm (quark)}$    & Fig. 2(a) &
       $(t,{\bar t}; b)$,  $(b,{\bar b}; t)$ \\
\hline\hline
\end{tabular}
\end{center}
\pagebreak
\begin{center}
{\bf Figure Captions}
\end{center}
\begin{description}
\item{Fig. 1}
Three types of Feynman diagrams contributing to 
the  two-point functions $\Pi _{ij}$, where $i$ and 
$j$ are either of $H$, $h$, $G$  or $A$. Fig. 1 (c) 
denotes the tad pole contributions. Internal particles 
$(X, Y)$ in (a) and $X$ in (b) are listed in Table 2.
\item{Fig. 2}
Three types of Feynman diagrams contributing 
to the $Hzz$ vertex;  
(a) $\Gamma _{Hzz}^{(1)}$ and $\Gamma _{Hzz}
^{\rm (quark)}$, (b) $\Gamma _{Hzz}^{(2)}$, and 
(c) $\Gamma _{Hzz}^{(3)}$. Internal particles $(X, Y; Z)$ 
in (a)  and $(X,Y)$ in (b) and (c) 
are given in Table 3.
\item{Fig. 3 }
The decay width 
$\Gamma (H \rightarrow Z_{L}Z_{L})$ ((a), (b) and (c))
and 
$\Gamma (H \rightarrow W_{L}^{+}W_{L}^{-})$ 
((A), (B) and (C))
as a function of $m_{G}$. The choice of 
$m_{A}$ is varied as 400 GeV ((a) and (A)), 
600 GeV ((b) and (B))
 and 800 GeV ((c) and (C)).
 Mixing angles 
are determined by $\tan \beta=2$, and 
$\sin ^{2}(\alpha -\beta)=1$, 
and the neutral $CP$-even Higgs boson masses are 
$m_{H}=300$ GeV and $m_{h}=400$ GeV. 

\item{Fig. 4 }
The decay width 
$\Gamma (H \rightarrow Z_{L}Z_{L})$ ((a), (b) and (c))
and 
$\Gamma (H \rightarrow W_{L}^{+}W_{L}^{-})$ 
((A), (B) and (C))
as a function of $m_{G}$. The choice of 
$m_{A}$ is varied as 400 GeV ((a) and (A)),
 600 GeV ((b) and (B))
 and 800 GeV ((c) and (C)).
 Mixing angles 
are determined by $\tan \beta=10$, and 
$\sin ^{2}(\alpha -\beta)=0.5$, 
and the neutral $CP$-even Higgs boson masses are 
$m_{H}=300$ GeV and $m_{h}=400$ GeV. 

\item{Fig. 5}
The bird-eye's view of the decay width $\Gamma 
(H\rightarrow Z_{L}Z_{L})$ as a function of 
$m_{G}$ and $m_{A}$. The mixing parameters are
determined by $\tan \beta=2$, and 
$\sin ^{2}(\alpha -\beta)=1$,
and the neutral $CP$-even Higgs boson masses are 
$m_{H}=300$ GeV and $m_{h}=400$ GeV. 
\item{Fig. 6}
The bird-eye's view of the decay width $\Gamma 
(H\rightarrow Z_{L}Z_{L})$ as a function of 
$m_{G}$ and $m_{A}$. The mixing parameters are
determined by $\tan \beta=10$, and 
$\sin ^{2}(\alpha -\beta)=0.5$,
and the neutral $CP$-even Higgs boson masses are 
$m_{H}=300$ GeV and $m_{h}=400$ GeV. 
\item{Fig. 7}
The bird-eye's view of the decay width $\Gamma 
(H\rightarrow W_{L}^{+}W_{L}^{-})$ as a function of 
$m_{G}$ and $m_{A}$. The mixing parameters are
determined by $\tan \beta=2$, and 
$\sin ^{2}(\alpha -\beta)=1$,
and the neutral $CP$-even Higgs boson masses are 
$m_{H}=300$ GeV and $m_{h}=400$ GeV. 
\item{Fig. 8}
The bird-eye's view of the decay width $\Gamma 
(H\rightarrow W_{L}^{+}W_{L}^{-})$ as a function of 
$m_{G}$ and $m_{A}$. The mixing parameters are
determined by $\tan \beta=10$, and 
$\sin ^{2}(\alpha -\beta)=0.5$,
and the neutral $CP$-even Higgs boson masses are 
$m_{H}=300$ GeV and $m_{h}=400$ GeV. 
\item{Fig. 9}
The mixing angle $\alpha $ dependence of the 
decay width $\Gamma (H\rightarrow Z_{L}Z_{L})$. 
The mixing angle $\beta$ is determined by 
(a) $\tan \beta =2$, (b) $\tan \beta =5$ and 
(c) $\tan \beta =10$, respectively.
The Higgs boson masses are 
$m_{H}=300$ GeV, $m_{h}=400$ GeV, $m_{G}=500$ GeV
 and $m_{A}=600$ GeV.
\end{description}
\end{document}